\documentclass[12pt]{article}
\usepackage{amsmath}
\usepackage{graphicx}
\usepackage{natbib}
\usepackage{url} 
\usepackage{graphicx}
\usepackage{amsmath}
\usepackage{natbib}
\usepackage{amscd,amssymb,amsfonts,verbatim}
\usepackage{mathrsfs}
\usepackage{bbm}
\usepackage{listings}
\usepackage{epsfig}
\usepackage{enumitem}
\usepackage{url}
\usepackage{booktabs,makecell,ltablex}
\usepackage{multirow}
\usepackage{tikz}
\usetikzlibrary{patterns}
\usetikzlibrary{trees}
\usepackage{pgfplots}
\usepgfplotslibrary{fillbetween}
\usetikzlibrary{shapes.misc}
\usetikzlibrary{shapes.geometric}

\pgfplotsset{compat=1.16}
\pgfplotsset{soldot/.style={color=blue,only marks,mark=*}}
\pgfplotsset{holdot/.style={color=blue,fill=white,only marks,mark=*}}
\usetikzlibrary{math}
\usetikzlibrary{patterns}
\usetikzlibrary{svg.path}
\def\bO{\mathbf{O}}
\def\bX{\mathbf{X}}
\def\bo{\mathbf{o}}

\def\bB{\mathbf{B}}

\def\q{\mathsf{q}}
\def\qone{\mathsf{q}_1}
\def\qtwo{\mathsf{q}_2}
\graphicspath{{./Figs/}}
\newcommand{\blind}{0}

\begin{document}

\def\spacingset#1{\renewcommand{\baselinestretch}%
{#1}\small\normalsize} \spacingset{1}


\if0\blind
{
  \title{\bf Bayesian inference for continuous-time hidden Markov models with an unknown number of states}
  \author{Yu Luo\thanks{
  		Department of Mathematics, Imperial College London, United Kingdom} ,  \hspace{.2cm}
  	David A. Stephens\thanks{
  		Department of Mathematics and Statistics, McGill University, Canada} \hspace{.2cm}
  }
 \date{ }
  \maketitle
} \fi

\if1\blind
{
  \bigskip
  \bigskip
  \bigskip
  \begin{center}
    {\LARGE\bf Bayesian inference for continuous-time hidden Markov models with an unknown number of states}
\end{center}
  \medskip
} \fi

\bigskip
\begin{abstract}
We consider the modeling of data generated by a latent continuous-time Markov jump process with a state space of finite but unknown dimensions. Typically in such models, the number of states has to be pre-specified, and Bayesian inference for a fixed number of states has not been studied until recently. In addition, although approaches to address the problem for discrete-time models have been developed, no method has been successfully implemented for the continuous-time case.  We focus on reversible jump Markov chain Monte Carlo which allows the trans-dimensional move among different numbers of states in order to perform Bayesian inference for the unknown number of states. Specifically, we propose an efficient split-combine move which can facilitate the  exploration of the parameter space, and demonstrate that it can be implemented effectively at scale. Subsequently, we extend this algorithm to the context of model-based clustering, allowing numbers of states and clusters both determined during the analysis. The model formulation, inference methodology, and associated algorithm are illustrated by simulation studies. Finally, We apply this method to real data from a Canadian healthcare system in Quebec.
\end{abstract}

\noindent%
\vfill

\newpage
\spacingset{1.5} 
\section{Introduction}
\label{sec:intro}
Continuous-time Markov processes on a finite state space have been widely used as models for irregularly spaced longitudinal data as they correspond to plausible data generating representations.  In almost all cases, the process is observed only at a number of discrete time points, rather than being continuously observed.  This problem that arises in a broad collection of practical settings from public health surveillance to molecular dynamics.  For example, healthcare systems and electronic health records represent large volumes of data that allow the calculation of longitudinal health trajectories; however, such health records are recorded only when patients interact with the health system.   Likelihood-based inference for the infinitesimal generator of a continuous-time Markov jump process has been detailed, for example, in \cite{jacobsen2012statistical}. However, in settings such as those identified above, inference for the infinitesimal generator becomes more difficult. \cite{bladt2005statistical} investigated likelihood-based inference for discretely observed continuous-time Markov processes, while \cite{tancredi2019approximate} proposed approximate Bayesian methods to facilitate the computation for such models.

In a related class of problems, the observed data are not directly representative of the Markov process, or similarly the process is observed with measurement error. In those cases, a hidden Markov model (HMM) is more appropriate: this model assumes that an unobserved stochastic process governs the generating model for observations, and assumptions of the Markov property are imposed on the unobserved sequence, with observations usually modeled as independent conditional on the hidden Markov process.  There is a broad interest in the application of the continuous-time HMM (CTHMM) in recent years, such as in ecological studies \citep{mews2020continuous} and in medical research \citep{lange2015joint,alaa2018hidden,lange2018estimating}, with a predominant focus on frequentist approaches. More recently, \cite{williams2019bayesian} and \cite{luocthmm2018a} implemented a fully Bayesian CTHMM using different missing data likelihood formulations for the underlying Markov chain.  Even when these models have been proposed and implemented, the number of states has had to be pre-specified. Determining the number of hidden states is a challenge addressed in earlier work \citep[see for example][]{celeux2008selecting,pohle2017selecting}.  \cite{luocthmm2018a} suggested using the BIC to select the number of states via the expectation–maximization algorithm before performing Bayesian inference with a fixed number of states. \cite{luo2019bayesian} extended Bayesian CTHMMs for finite and Dirichlet mixture model-based clustering procedures to cluster individuals, which allows Markov chain Monte Carlo (MCMC) to change the dimension of the number of clusters,  but still relied on the assumption that the number of states has to be pre-specified.

Bayesian model determination approaches have been a longstanding focus of interest in Bayesian inference \citep[see, for example,][]{carlin1995bayesian,green1995reversible,godsill2001relationship}. In particular, reversible jump MCMC \citep{green1995reversible} has provided a general solution by exploiting trans-dimensional moves that exploit the dynamics of the Metropolis-Hastings (MH) algorithm in a fixed dimension, allowing movement across parameter spaces of different dimensions. \cite{richardson1997bayesian} developed a reversible jump MCMC approach to univariate Normal mixture models, and subsequently \cite{robert2000bayesian} extended this work to discrete-time hidden Markov models with Normal mixtures. In their work, they specifically used two types of reversible jump moves in MCMC to explore the parameter space, i.e., split--combine and birth--death moves. \cite{stephens2000bayesian} introduced an alternative MCMC approach, using a birth--death point process to infer the number of components in the Normal mixture model setting, and \cite{cappe2003reversible} demonstrated the limit-case equivalence of the reversible jump and continuous-time methodologies for both mixture models and discrete-time HMMs. In this paper, we focus on constructing reversible jump MCMC for CTHMMs which allow the number of hidden states to be inferred via the posterior distribution.

For a better understanding of dynamic changes of individual trajectories, it would be helpful to cluster individuals trajectories and to study the pattern in each group to explore the variation in trajectories. Many of these methods may be classified as model-based clustering procedures, where clustering is achieved by consideration of parametric likelihood- or density-based calculations, with the number of clusters determined by information criteria, such as AIC or BIC \citep{dasgupta1998detecting,fraley1998many}. Similarly, however, in such calculations, the number of clusters has to be fixed, and determining the number of clusters is a challenge addressed by many clustering algorithms. We address this problem subsequently by extending our reversible jump MCMC procedures to allow the number of clusters to be inferred during the analysis.

The rest of the paper is organized as follows. In Section \ref{cthmm}, we describe the CTHMM-GLM. Section \ref{rjmcmc} presents fully Bayesian inference via reversible jump MCMC, specifically a split-combine move to update the number of states and then fix dimensional MCMC. Section \ref{cluster} extends the reversible jump MCMC approach to model-based clustering, allowing numbers of states and clusters to vary simultaneously. Simulation examples to examine the performance of proposed MCMC are presented in Section \ref{sec:sim}. Finally, we present the results from applying this method to a  chronic obstructive pulmonary disease (COPD) cohort in Section \ref{real}, and discuss these results in Section \ref{dis}.

\section{A Continuous-Time Hidden Markov Model}
\label{cthmm}
We presume that the data $\{O_1,\ldots,O_T\}$ recorded at observation time points $\{\tau_1,\dots,\tau_T\}$ arise as a consequence of a latent continuous-time Markov chain (CTMC) $\{X_s, s \in \mathbb{R}^+\}$ taking values on the finite integer set $\{1,2,\ldots,K\}$. Note that observations are indexed using an integer index (that is, $O_t$), and that the latent process is indexed using a continuous-valued index (that is, $X_{\tau_t}$).

Conditional on the latent process, we assume the observations are drawn from an exponential family model with density $f\left( {O_t} | X_{\tau_t} = k \right)$. If there are time-varying explanatory variables $\mathbf{Z}\in\mathbb{R}^D$, a generalized linear model (GLM) with link function $g(.)$ and regression coefficients $\beta_{k}$ for state $k$, is adopted. Define matrix $\bB = (\beta_{d,k})$ for $d=1,\ldots,D$ and $k =1,\ldots,K$ as the GLM coefficient matrix.  Finally, let $S_t=\left(S_{t,1},\ldots,S_{t,K}\right)^\top$ be an indicator random vector with $S_{t,k}=1$ if $X_{\tau_t}=k$ and 0 otherwise.

In the assumed model the data generating mechanism is specified via (i) the latent state model $X_s\left| \Theta  \right. \sim \text{CTMC}\left(\pi ,Q \right)$, where $Q$ is the infinitesimal generator and $\pi$ is the initial distribution for the continuous-time Markov process,  and (ii) the observation model $O_{\tau_t }\left| X_{\tau_t} = k \right. \sim{\text{GLM}}\left( \beta_{d,k} \right), d=1,\ldots,D$.  The model is parameterized by $\Theta=\left\{Q,\pi,\bB\right\}$; recall that the structural constraint on $Q$ is that its off-diagonal elements $\{q_{j,k}, j,k=1,\ldots,K, j \neq k \}$ are positive, and that its rows sum to zero.  In this paper, we impose no other constraints, although to do so would be straightforward: for example, we might wish to restrict certain $q_{j,k}$ to obey with linear constraints such as equality to zero.  In the model, the observations $\{O_1,\ldots,O_T\}$ are assumed mutually conditionally independent given $\{X_s\}$; this assumption is common, but can be easily relaxed.  With the Markov chain observed discretely at different time points, one could compute the likelihood function for $Q$ in \cite{jackson2003multistate,williams2019bayesian}, however to facilitate the MCMC algorithm, we consider the complete but unobserved trajectory of $\{X_s\}$ as a collection of auxiliary variables in a missing data formulation: the unobserved trajectory comprises a collection of states and transition times that completely describe the latent path over any finite interval. The detailed derivation of the  complete data likelihood, $\mathcal{L} (\Theta )$, is given in the Supplement.

Bayesian inference for this model with the number of states $K$ fixed has been fully studied by \cite{luocthmm2018a}, where an MCMC scheme based on simulating the complete latent path for each individual is developed; this MCMC scheme relies upon the rejection sampling approach of \cite{hobolth2009simulation} to sample the latent paths in an efficient fashion. Bayesian inference using the complete data likelihood formulation is appealing as it produces posterior samples of the full unobserved state sequences and latent continuous-time process, which allows inference to be made for individual-level trajectories across the entire observation window, and which is useful for computing posterior distributions for pathwise aggregate features on individual trajectories.

\section{Reversible Jump MCMC for CTHMMs}
\label{sec:rjmcmc}
First, we add the number of states $K$ as an additional parameter and extend the MCMC algorithm to allow for inference to be made via the posterior distribution for $K$.  There are several different approaches that can be adopted that we outline below and in the Supplement.  First we study split/combine moves for states/pairs of states similar in spirit to the split/merge moves of \cite{richardson1997bayesian,dellaportas2006multivariate}. The Supplement also gives detailed descriptions and simulation examples for inference on $K$ via a birth-death point process by \cite{stephens2000bayesian}.  In our analysis, we restrict attention to the case where transition dynamics are modeled homogeneously across the sample, that is, $q_{n,l,m} \equiv q_{l,m}$ for all $n,l,m$.

\label{rjmcmc}
\subsection{Markov chain Monte Carlo methodology}
One iteration of the MCMC algorithm that incorporates the required trans-dimensional move proceeds using the following two types of move:
\begin{enumerate}
	\item A split/combine move that considers splitting a hidden state into two, or combining two hidden states into one.
	\item With the number of states $K$ fixed, update the model parameters using standard MCMC moves:
	\begin{itemize}
		\item update latent state indicators $S_{n,t}$;
		\item update the parameters associated with the observation process $\bB$;
		\item update the initial distribution $\pi$;
		\item update the infinitesimal generator $Q$.
	\end{itemize}
\end{enumerate}
The Supplement gives detailed procedures of updating the model parameters with a fixed number of states, which was extensively studied in \cite{luocthmm2018a}. Specifically, for split and combine moves, we will implement the reversible jump algorithm by \cite{green1995reversible}. Consider a proposal from the current model state $(K,\Theta)$ to a new state $(K',\Theta')$ using the proposal density
\[
\q\left(K',\Theta';K,\Theta\right) = \qone\left(K';K\right)\qtwo\left(\Theta_{K'};\Theta_K\right)
\]
that is, using independent proposals for the two components. The acceptance probability for this form of proposal using the MH procedure is given by
\begin{equation*}
	\begin{aligned}
		\alpha\left(K',\Theta';K,\Theta\right) &= \min \left(1, \frac{\q\left(K,\Theta;K',\Theta'\right)p\left(K',\Theta'\left|\bo\right.\right)}{\q\left(K',\Theta';K,\Theta\right)p\left(K,\Theta\left|\bo\right.\right)}\right)\\[6pt]
		&=\min \left(1, \frac{\qone\left(K;K'\right)\qtwo\left(\Theta_{K};\Theta_{K'}\right)p\left(K',\Theta'_{K'}\left|\bo\right.\right)}{\qone\left(K';K\right)\qtwo\left(\Theta_{K'};\Theta_K\right)p\left(K,\Theta_K\left|\bo\right.\right)}\right)\\
	\end{aligned}
\end{equation*}
where $p\left(K,\Theta_K\left|\bo\right.\right)$ is the posterior distribution of $(K,\Theta_K)$ given the observed data $\bo$, which can, up to proportionality, be decomposed into the marginal (or `incomplete data') likelihood of the data $\mathcal{L} (\bo| \Theta_K,K )$ times the prior distribution for $(K,\Theta_K)$;
\[
p\left(K,\Theta_K\left|\bo\right.\right) \propto \mathcal{L} (\bo| \Theta_K,K ) p_0(\Theta_K|K) p_0(K).
\]
Our algorithm relies upon the ability to compute the marginal likelihood efficiently for any $\Theta_K$; however, this is a standard `forward' calculation for CTHMMs.

\subsection{Split and Combine Moves}
\label{sec:smstates}
To construct efficient split and combine moves under the reversible jump framework, we adopt the idea of centered proposals by \cite{brooks2003efficient}. The proposal is designed to produce similar likelihood contributions for the current and proposed parameters. The combine move is designed to choose a state, $k$ at random and select another state $k'$ such that $\sum_{i=1}^{D}\left|\beta_{k,i}-\beta_{j,i}\right|$ is smallest for $j\ne k$. The reverse split move is to randomly select a cluster, $k$  to split into two clusters, say $k$ and $k'$, and check if the condition, $\sum_{i=1}^{D}\left|\beta_{k,i}-\beta_{k',i}\right| <\sum_{i=1}^{D}\left|\beta_{k,i}-\beta_{j,i}\right|$ for $j\ne k,k'$. If this condition is not met, then the split move is rejected right away.
\subsubsection{Split Move}
We consider an update that changes $K \to K+1$, requiring the generation of a new hidden state.  For this move, we set $\qone\left(K;K+1\right) = \qone\left(K+1;K\right)$ for each $K$. Then the acceptance probability reduces to
\begin{equation}
	\label{MHa}
	\alpha\left(K+1,\Theta_{K+1};K,\Theta_K\right) =\min \left(1, \frac{\qtwo\left(\Theta_{K};\Theta_{K+1}\right)p\left(K+1,\Theta_{K+1}\left|\bo\right.\right)}{\qtwo\left(\Theta_{K+1};\Theta_K\right)p\left(K,\Theta_K\left|\bo\right.\right)}\right).
\end{equation}
We denote the posterior ratio in the final term $r\left(K+1,\Theta_{K+1};K,\Theta_K|\bo\right)$, that is
\[
r\left(K+1,\Theta_{K+1};K,\Theta_K|\bo\right) = \dfrac{p\left(K+1,\Theta_{K+1}\left|\bo\right.\right)}{p\left(K,\Theta_K\left|\bo\right.\right)}.
\]
First, we randomly select a state on which to perform the split move. Without loss of generality, we consider the case where state $K$ is to be split into new states $K$ and $K+1$. We propose the new $(K+1)$-dimensional infinitesimal generator $Q_{K+1}$ using the following updates:
\begin{equation}
	\begin{array}{ccc}
		q'_{K,j}=q_{K,j} & q'_{K+1,j}=q_{K,j}   & 1\le j <K\\[6pt]
		q'_{i,K}=w_{i}q_{i,K} & q'_{i,K+1}=\left(1-w_{i}\right)q_{i,K} & 1\le i <K\\[6pt]
		w_{i} \sim Beta\left(2,2\right)&q'_{K,K+1}, q'_{K+1,K} \sim p_{0Q}
	\end{array}
\end{equation}
with $Q_K= \left\{q_{i,j}\right\}_{1\le i,j \le K}$ from the original $K$-state model  and  $Q_{K+1}=\{q'_{i,j}\}_{1\le i,j \le K+1}$; here, $p_{0Q}\left(.\right)$ is the prior distribution for the element in $Q$, which is assumed to be $Gamma\left(a,b\right)$. In this way, the new stationary probabilities $s'$ of the CTMC associated with $Q_{K+1}$ satisfying $s'Q_{K+1}=0$ are $s'_j=s_j$ for $1 \le j <K$, $s_K=s'_K+s'_{K+1}$ where $s$ is a vector of stationary probabilities associated with $Q_K$ (satisfying $sQ_{K}=0$). The dynamical properties of the CTMC are thus preserved.  The observation process parameters associated with new state $K+1$ are generated as
\[
\beta'_{1,K+1} \sim \mathcal{N}\left(\beta_{1,K},c^2\right), \;\;\;
\beta'_{m,K+1}=\beta_{m,K}, \;\; 2\le m \le D
\]
and the remaining elements of $\bB_{K+1}$ set equal to the elements of $\bB_{K}$.  In addition, we generate a weight $w \sim Beta\left(2,2\right)$ to split the initial probability for state $K$ in  $\pi^K=\left(\pi_1,\ldots,\pi_K\right)^\top$ into $\pi'_{K}=w\pi_K$ and $\pi'_{K+1}=(1-w)\pi_{K}$ and the rest remains the same.  In the acceptance probability in \eqref{MHa}, the ratio of the proposal density can be written as
\[
\frac{\qtwo\left(\Theta_{K};\Theta_{K+1}\right)}{\qtwo\left(\Theta_{K+1};\Theta_K\right)}=\frac{\q\left(Q_K;Q_{K+1}\right)}{\q\left(Q_{K+1};Q_{K}\right)}\times \frac{\q\left(\bB_K;\bB_{K+1}\right)}{\q\left(\bB_{K+1};\bB_{K}\right)}\times \frac{\q\left(\pi^K;\pi^{K+1}\right)}{\q\left(\pi^{K+1};\pi^{K}\right)}.
\]
Specifically,
\[
\frac{\q\left(Q_K;Q_{K+1}\right)}{\q\left(Q_{K+1};Q_{K}\right)}=\frac{\prod\limits_{i=1}^{K-1}q_{i,K}}{ p_{0Q} (q'_{K,K+1} )p_{0Q} (q'_{K+1,K} )\prod\limits_{i=1}^{K-1}p\left(w_i\right)}
\]
where the numerator comes from the Jacobian of the transformation that creates the proposed $Q_{K+1}$. Then
\[
\frac{\q\left(\bB_K;\bB_{K+1}\right)}{\q\left(\bB_{K+1};\bB_{K}\right)}=\frac{1}{p (\beta'_{1,K} )}
\]
where $p (\beta'_{1,K} )$ is the Normal density with mean $\beta_{1,K}$ and variance $c^2$.
\[
\frac{\q\left(\pi^K;\pi^{K+1}\right)}{\q\left(\pi^{K+1};\pi^{K}\right)}=\frac{\pi_{K}}{p\left(w\right)}
\]
where the numerator comes from the Jacobian of the transformation that generates $\pi^\prime$. Therefore the MH acceptance probability with the prior distribution as $p_0$ is
\begin{equation}
	\label{as}
	\begin{aligned}
		&\alpha\left(K+1,\Theta_{K+1};K,\Theta_K\right) \\
		=& \min \left(1, \frac{\q\left(Q_K;Q_{K+1}\right)}{\q\left(Q_{K+1};Q_{K}\right)} \frac{\q\left(\bB_K;\bB_{K+1}\right)}{\q\left(\bB_{K+1};\bB_{K}\right)}  \frac{\q\left(\pi^K;\pi^{K+1}\right)}{\q\left(\pi^{K+1};\pi^{K}\right)} r\left(K+1,\Theta_{K+1};K,\Theta_K|\bo\right) \right)\\[6pt]
		=&\min \left(1, \frac{d_{K+1}\times\prod\limits_{i=1}^{K-1}q_{i,K} \times \pi_{K} }{b_K p_{0Q}  (q'_{K,K+1} )p_{0Q} (q'_{K+1,K} )\prod\limits_{i=1}^{K-1}p (w_i )p (\beta'_{1,K} ) p\left(w\right) } r\left(K+1,\Theta_{K+1};K,\Theta_K|\bo\right) \right)\\
	\end{aligned}
\end{equation}
where $b_K$ is the probability of choosing the split move and $d_{K+1}=1- b_K$ is the probability of choosing the combine move.
\subsubsection{Combine Move}
For the update from $K+1$ to $K$ states, we consider the following move.  Without loss of generality, we consider how to combine states $K$ and $K+1$ into a single new state $K$. For the current configuration $Q_{K+1}$, we propose the move to $Q_{K}$ as
\begin{equation*}
	\begin{array}{clc}
		q_{K,j}& = \dfrac{s'_K}{s'_K+s'_{K+1}}q'_{K,j}+  \dfrac{s'_{K+1}}{s'_K+s'_{K+1}} q'_{K+1,j}   & 1\le j <K\\[12pt]
		q_{i,K}& = q'_{i,K}+q'_{i,K+1} & 1\le i <K
	\end{array}.
\end{equation*}
The remaining $q_{ij}$, where $i\ne j$, are obtained by copying  $Q_{K+1}$ and discarding $q'_{K,K+1}$ and $q'_{K+1,K}$, with the diagonal terms of $Q_{K}$  calculated by $q_{ii} = -\sum\limits_{j\ne i} q_{ij}$ for $1\le i \le K$. It can be verified that the stationary probabilities, $s=\left(s_1,\ldots,s_K\right)^\top$ associated with $Q_K$, are $s_j=s'_j$ for $1\le j <K$ and $s_K=s'_K+s'_{K+1}$.  For the parameters in observation process, we propose
\[
\beta_{m,K}=\frac{s'_K}{s'_K+s'_{K+1}}\beta'_{m,K}+\frac{s'_{K+1}}{s'_K+s'_{K+1}}\beta'_{m,K+1} \;\;\;\; 1\le m\le D.
\]
The remaining elements of $\beta_{m,j}$ for $j<K$ are taken to be the same as the current parameter configuration $\bB_{K+1}$. Finally, we propose the initially distribution $\pi^{K+1}=\left(\pi'_1,\ldots,\pi'_K,\pi'_{K+1}\right)$ simply moves to $\pi^{K}=\left(\pi_1,\ldots,\pi_K\right)$  where $\pi_K=\pi'_K+\pi'_{K+1}$ and $\pi_j=\pi'_j$ for $j<K$. Therefore, the proposal ratio is computed as follows:
\[
\frac{\q\left(\pi^{K+1};\pi^{K}\right)}{\q\left(\pi^K;\pi^{K+1}\right)}=\frac{p\left(w\right)}{\pi_{K}}.
\]
This is the reverse move corresponding to the split move described above, and essentially $w= \pi'_{K}/\pi_{K}$ and $p(.)$ is the density of $Beta(2,2)$. For the infinitesimal generator, the reverse move for $q_{i,K}$ for $1 \le i <K$ is the same with the split move. The reverse move for  $q_{K,j}$, $1 \le j <K$, can be viewed as
\[
q'_{K,j}=\frac{u_1}{u_0}q_{K,j}\;\;\;\;\; q'_{K+1,j}=\frac{1-u_1}{1-u_0}q_{K,j}
\]
where $u_0=s'_K/(s'_K+s'_{K+1})$ and $u_1$ is a weight parameter. If we choose $u_0=u_1$, then $q'_{K,j}=q_{K,j}$ and $q'_{K+1,j}=q_{K,j}$. This reverses what was proposed for the split move. Therefore,
\[
\frac{\q\left(Q_{K+1};Q_{K}\right)}{\q\left(Q_K;Q_{K+1}\right)}=\frac{p_{0Q} (q'_{K,K+1} ) p_{0Q}  (q'_{K+1,K} )p\left(w_i\right)}{\prod\limits_{i=1}^{K-1}q_{i,K}}.
\]
In terms of $B$, since we mimicked the proposal for $q_{K,j}$, therefore the reverse move is $\beta'_{m,K}=\beta_{m,K}$ and $\beta'_{m,K+1}=\beta_{m,K}$ for $1\le m \le D$. Then the proposal ratio $\q\left(\bB_{K+1};\bB_{K}\right)/\q\left(\bB_K;\bB_{K+1}\right)$ equals 1, and the MH acceptance probability for the combine move, $\alpha\left(K,\Theta_K;K+1,\Theta_{K+1}\right)$, is the minimum of 1 and
\begin{equation}
	\label{am}
	\frac{b_Kp_{0Q} (q'_{K,K+1} )p_{0Q} (q'_{K+1,K} )\prod\limits_{i=1}^{K-1}p\left(w_i\right) p\left(w\right) }{d_{K+1} \prod\limits_{i=1}^{K-1}q_{i,K} \pi_{K}}  r\left( K,\Theta_{K};K+1,\Theta_{K+1}|\bo \right).
\end{equation}

\section{Model-Based Clustering for CTHMMs}
\label{cluster}
So far, we have constructed fully Bayesian inference for a CTHMM via reversible jump MCMC, allowing the number of states to vary during the analysis. We now extend this methodology to cluster trajectories based on a CTHMM with an unknown number of states.  Specifically, we will employ model-based clustering procedures to cluster individuals based on the component model parameters that determine the mixture form. The basic formulation of the model envisages that the population is composed of distinct sub-populations each with a distinct stochastic property. For a CTHMM, this corresponds to each group having a potentially different component of parameter $\Theta = (\pi, Q, \bB)$ and the number of states, $K$.  \cite{luo2019bayesian} develop model-based clustering for CTHMMs under finite and infinite mixture models, with a fixed number of states. We incorporate this finite mixture model structure into the proposed reversible jump MCMC, allowing both the number of states and the number of clusters to be inferred during the analysis. There is a crucial distinction between the number of components $M$ in the mixture model and the number of clusters $M^*$ in the data which is defined as the number of components used to generate the observed data, or the number of ``filled" mixture components. In the algorithm described below, we focus on specifying a prior on the number of components $M$, which implicitly places a prior on $M^*$ \citep{miller2018mixture}; however, in our simulation and real examples, the proposed split move merely generates any empty component. For a comprehensive investigation of  $M$ and $M^*$ in different trans-dimensional algorithms, see \cite{fruhwirth2020generalized}. 

Let $M$ be the number of components and $C_n$ be the cluster membership indicator for individual $n$.  For $n=1,2,\ldots,N$, it is presumed to be a member of a component labelled $1,2,\ldots,M$, where $\varpi_m = \mathbb{P}\left( {C_n = m} \right)$ is the prior probability that individual $n$ is assigned to component $m$, subject to $\sum\limits_{m=1}^M  \varpi_M=1$.  The following hierarchy leads to model-based clustering procedures for the CTHMM:
\begin{align*}
	M &\sim p_0\left(M\right),  \text{a mass function on} \left\{1,2,3,\ldots\right\}\\
	\varpi_1,\ldots,\varpi_M\left|M\right. &\sim \text{Dirichlet}\left(\delta,\ldots,\delta\right)\\
	\mathbb{P}\left(C_n=m\left|\varpi_1,\ldots,\varpi_M, M\right.\right)& =\varpi_m, m=1,\ldots, M; n=1.\ldots N \\
	K_m &\sim p_0\left(K\right),  \text{a mass function on} \left\{1,2,3,\ldots\right\}, m=1,\ldots, M\\
	X_{n} \left| \Theta, C_n, K_{C_n} \right. &\sim \text{CTMC}\left(\pi^{(C_n)} ,Q^{(C_n)}\right)\\
	O_n \left| {X_{n},\Theta,C_n, K_{C_n}} \right. & \sim{\text{Exponential Family}}\left( {{B^{(C_n)}}} \right)
\end{align*} 
with $\delta=1$, making the weight distribution uniform. Then the complete-data likelihood for subject $n$ is
\[
\mathcal{L}\left(C_n,O_n,X_n,\Theta\right)=\prod \limits_{m=1}^M {\left[ \varpi_m \mathcal{L}\left(O_n,X_n \left|C_n=m,\Theta^{(m)} ,K_{m} \right. \right)\right]^{\mathbbm{1}\left(C_n=m\right)}}
\]
where $\mathbbm{1}\left(C_n=m\right)$ is the indicator function. A subject is assigned to component $m$ according the posterior probability
\begin{equation}
	\label{postfin}
	\mathbb{P}\left(C_n=m\left|{O_n,X_n},\Theta, K_{m} \right.\right)=\frac{{{\varpi_m}\mathcal{L}\left( {O_n,X_n\left| {C_n = m},\Theta^{(m)}, K_{m} \right.} \right)}}{{\sum\limits_{l = 1}^M {{\varpi_l}\mathcal{L}\left( {O_n,X_n\left| {C_n = l},\Theta^{(l)},K_{l} \right.} \right)} }}.
\end{equation}
In reality, the model parameter, $\Theta$, and the values of the latent states, $X_n$, are not known, and must be inferred from the observed data.

\subsection{Reversible-Jump MCMC for Clustering with An Unknown Number of States}
In \cite{luo2019bayesian}, a reversible-jump algorithm based on the marginalized model in \eqref{marg} was used to update $M$, and we will incorporate this move into our algorithm in Section \ref{sec:rjmcmc} to construct a clustering mechanism which allows the number of clusters and the number of states determined together during the analysis. We first apply the reversible-jump MCMC algorithm in Section \ref{sec:rjmcmc} to update the number of states in each component. We then update the number of components according to a split-combine move, while the combine move only involves components with the same number of states. We summarize one iteration of this clustering mechanism as follows: 

\begin{enumerate}
	\item Update the number of states for each component using the algorithm in Section \ref{sec:smstates}; If the move is accepted, update the model parameter in the corresponding component.
	\item Update the number of components by splitting a component or combining  components with the same number of states; If a component with $K_m$ states is chosen in the split move, then the move is to consider splitting the component into two both with $K_m$ states; If two components with the same number of states, $K_m$, are selected in the combine move, then the move is to combine  two components into one component with $K_m$ states.
	\item Given parameters in each component, update the component membership for each individual according to the posterior probability \eqref{postfin}.
	\item With the number of components $M$ fixed, each with fixed states $K_m$ where $m=1,\ldots,M$, update the model parameters using standard MCMC moves in each component, which the detail is given in the Supplement.
\end{enumerate}

For any empty component from Step 3, we generate model parameters from prior distributions. For the split and combine moves in (b),  we carry out them on the marginalized model as in \cite{luo2019bayesian}, where component labels and latent processes are marginalized out from the calculation, and use the likelihood
\begin{equation}
	\label{marg}
	\mathcal{L} (\bo|\Theta ,M ) = \prod_{n=1}^N \left\{ \sum_{m=1}^M \varpi_m \mathcal{L}(O_n|\Theta^{(m)},K_m ) \right\}.
\end{equation}
Similar with updating the number of states, we update $M$ by considering a proposal from the current state $(M,\Theta)$ to a new state $(M',\Theta')$ using the proposal density
$\q\left(M',\Theta';M,\Theta\right) = \qone\left(M';M\right)\qtwo\left(\Theta';\Theta\right)$, that is, using independent proposals for the two components. The acceptance probability for this proposal is given by
\begin{equation*}
	\begin{aligned}
		\alpha\left(M',\Theta';M,\Theta\right)
		&=\min \left(1, \frac{\qone\left(M;M'\right)\qtwo\left(\Theta ;\Theta'\right) p\left(M',\Theta'\left|\bo\right.\right)}{\qone\left(M';M\right)\qtwo\left(\Theta';\Theta \right)p\left(M,\Theta \left|\bo\right.\right)}\right)\\
	\end{aligned}
\end{equation*}
where $p\left(M,\Theta\left|\bo\right.\right)$ is the posterior distribution of $(M,\Theta)$ given the observed data $\bo$, which can, up to proportionality, be decomposed into the marginal likelihood of the data $\mathcal{L} (\bo| \Theta,M )$ times the prior distribution for $(M,\Theta)$, with prior distribution as $p_0$;
\[
p\left(M,\Theta\left|\bo\right.\right) \propto \mathcal{L} (\bo| \Theta,M ) p_0(\Theta|M) p_0(M).
\]
We will discuss trans-dimensional moves for updating the number of components in more detail below.
\subsection{Split/Combine Move for Updating the Number of Clusters}
The combine move is designed to choose a component, $m$ say, at random and select another component $i$ such that $\left\|\bB_i-\bB_m\right\|_2$ is smallest for $i\ne m$. The reverse split move is to randomly select a component, $m$  to split into two components, say $m$ and $m^*$, and check if the condition, $\left\|\bB_{m*}-\bB_m\right\|_2 <\left\|\bB_j-\bB_m\right\|_2$ for $j\ne m$. If this condition is not met, then the split move is rejected.

\subsubsection{Split Move}
We consider an update that changes the number of component from $M \to M+1$. Without loss of generality, we aim to split the $M^\text{th}$ component with CTMC parameters $\Theta_{M}=\{\pi_{M},Q_{M},\bB_{M}\}$ into two components, requiring the need to generate $K_M$ new hidden states, with corresponding parameters, i.e., $\Theta{'}=\{\pi{'},Q{'},\bB{'}\}$ and $\Theta{''}=\{\pi{''},Q{''},\bB{''}\}$. To implement the idea of centering proposals, we use a deterministic proposal for $Q$ and $\pi$, and let $Q{'}=Q{''}=Q_{M}$ and $\pi^{'}=\pi^{''}=\pi_{M}$. For observation parameter $\bB$, we can use the similar proposal:
\[
\beta_{1,k}^{'}=\beta_{M,1,k} \quad \beta_{1,k}^{''}\sim \mathcal{N}\left(\beta_{M,1,k},c^2\right)  \qquad k=1,\ldots, K_M
\]
\[
\beta_{j,k}^{'}= \beta_{j,k}^{''}=\beta_{M,j,k},\;   \qquad j=2,\ldots, D.
\]
For  mixture weights $\varpi$, let $w\sim Beta(2,2)$ and define $\varpi^{'}=w  \varpi_{M}$ and  $\varpi^{''}=(1-w) \varpi_{M}$. If we define the posterior ratio as
\[
r_c\left(M+1,(\Theta^{'},\Theta^{''},\varpi',\varpi'');M,(\Theta_{M},\varpi_M)|\bo\right) = \dfrac{p\left(M+1,(\Theta^{'},\Theta^{''},\varpi',\varpi'')\left|\bo\right.\right)}{p\left(M,(\Theta_{M},\varpi_M)\left|\bo\right.\right)}.
\]
Then, the acceptance probability for this proposal is
\begin{equation}
	\label{ascl}
	\begin{aligned}
		\min \big(1,& \frac{\q\left(Q_{M};Q^{'},Q^{''}\right)}{\q\left(Q^{'},Q^{''};Q_{M}\right)} \frac{\q\left(\bB_{M};\bB^{'},\bB^{''}\right)}{\q\left(\bB^{'},\bB^{''};\bB_{M}\right)}  \frac{\q\left(\pi_{M};\pi^{'},\pi^{''}\right)}{\q\left(\pi^{'},\pi^{''};\pi_{M}\right)}
		\frac{\q\left(\varpi_{M};\varpi^{'},\varpi^{''}\right)}{\q\left(\varpi^{'},\varpi^{''};\varpi_{M}\right)}\\[6pt]
		&\times r_c\left(M+1,(\Theta^{'},\Theta^{''},\varpi',\varpi'');M,(\Theta_{M},\varpi_M)|\bo\right) \big)\\[6pt]
		= \min &\left(1, \dfrac{d_{M+1} \varpi_{M}}{b_M   p_{\varpi}(w )p_{\beta}(\beta_{1,k}^{''}) } r_c\left(M+1,(\Theta^{'},\Theta^{''},\varpi',\varpi'');M,(\Theta_{M},\varpi_M)|\bo\right) \right)
	\end{aligned}
\end{equation}
where $b_M$ is the probability of choosing the split move and $d_{M+1}=1- b_M$ is the probability of choosing the combine move, and $p_{\beta} (\cdot)$ is the Normal density with mean $\beta_{1,K}$ and variance $c^2$ and $p_{\varpi} (\cdot)$ is the $Beta(2,2)$ density. 

\subsubsection{Combine Move}
For the combine move, we need choose two components with the same number of states and update from $M+1 \to M$ components. Again, without  loss of generality, we consider combine the $(M+1)^\text{th}$ and $M^\text{th}$ components into one component, both components with $K_M=K_{M+1}$ states, with the proposed parameters, $\Theta{'}=\{\pi{'},Q{'},\bB{'}\}$. We first find the stationary probabilities, $s_{M}$ and $s_{M+1}$,  associated with $Q_{M}$ and  $Q_{M+1}$. To combine $Q_{M}$ and  $Q_{M+1}$ into $Q^{'}$, the operation is as follows:
\[
q_{i,k}^{'}=\frac{s_{M,i}}{s_{M,i}+s_{M+1,i}}\times q_{M,i,k}+\frac{s_{M+1,i}}{s_{M,i}+s_{M+1,i}}\times q_{M+1,i,k}, i\ne k=1,\ldots,K_M
\]
and $q_{M,k,k}=-\sum_{i \ne k}q_{M,i,k}$ for $k=1,\ldots,K_M$. For the observation process parameter $\bB$,
\[
\beta_{i,k}^{'}=\frac{s_{M,i}}{s_{M,i}+s_{M+1,i}}\times \beta_{M,i,k}+\frac{s_{M+1,i}}{s_{M,i}+s_{M+1,i}}\times \beta_{M+1,i,k}, i, k=1,\ldots,K_M.
\]
For the initial distribution $\pi$, 
\[
\pi_{k}^{'}=\frac{s_{M,i}}{s_{M,i}+s_{M+1,i}}\times \pi_{M,k}+\frac{s_{M+1,i}}{s_{M,i}+s_{M+1,i}}\times \pi_{M+1,k},  k=1,\ldots,K_M
\]
and rescale the sum to 1. For mixture weights $\varpi$, the move is essentially to add up the probability of the two corresponding components, i.e., $\varpi^{'}=\varpi_{M}+\varpi_{M+1}$. The acceptance probability from $M+1$ to $M$ components is
\begin{equation}
	\label{amcl}
	\min\left( 1,\frac{b_M p_{\varpi}(w ) }{d_{M+1}\varpi^{'}}  r_c\left(M,\Theta{'};M+1,\Theta_{M},\Theta_{M+1}|\bo \right)\right).
\end{equation}

\section{Simulation}
\label{sec:sim}
In this section we demonstrate the performance of the proposed
reversible jump and birth-death (In the Supplement) MCMC approaches for the CTHMM.

\subsection{Identifying the number of states}
\label{sim:rjmcmc}
In the first example, we demonstrate the performance of MCMC to estimate the number of states, and to discover how performance degrades when the problem becomes more challenging. We consider a four-state model with coefficients
\begin{equation*}
	Q=\left(\begin{array}{rrrr}
		-3.00 & 2.00&  1.00& 0.00\\
		1.00 & -1.80  &0.75 & 0.05\\
		0.15&  0.55 & -1.05 & 0.35 \\
		0.00 & 0.25 & 0.40& -0.65\\
	\end{array}
	\right)
\end{equation*}
and with two time-varying covariates $Z_1\sim \mathcal{N}\left(-1,1\right)$ and $Z_2\sim \text{Binomial}\left(1,0.6\right)$, with
\begin{equation*}
	\bB=\left(\begin{array}{rrrr}
		-1.28 & -0.55 & -1.05 & 0.99\\
		-0.88 & 1.15  &1.36 & 1.73\\
		0.70 &0.68 &-1.12 &-1.20
	\end{array}
	\right)
\end{equation*}
The initial distribution $\pi$ is set to be $\left(0.35,0.25,0.2,0.2\right)$. We first constructe the continuous time Markov process from the generator $Q$ for subject $i$, a continuous-time realization of the latent sequence $\left\{X_s,0 \le s \le 15\right\}$, and uniformly at random extract $T$  observation time points between 0 and 15, where $T \sim Uniform(20,60)$.  We assume that the first observation is made at time 0. We generate the data for 1000 subjects.  The prior distributions for the elements in $Q$ and $\pi$ are specified as independent $Gamma(1,2)$ and $Dirichlet(1,\ldots,1)$. Non-informative prior is imposed for $\bB$. We use a $Poisson(3.5)$ distribution as the prior for the number of states, and initiate the model with one hidden state.

\begin{table}
	\caption{\label{simex1} Example \ref{sim:rjmcmc}: Posterior distribution of the number of hidden states. The true number of states is four.}
	\centering
	\fbox{%
		\begin{tabular}{*{5}{c}}
			\# of hidden states & Normal $\sigma=1$ & Normal $\sigma=1.5$ & Normal $\sigma=2$ & Poisson\\
			\hline
			1& 0.0001 &0.0001& 0.0001&0.0001\\
			2 & 0.0005&0.0001& 0.0001 &0.0002\\
			3 & 0.0002&0.0002& 0.0002 &0.0001\\
			
			4 & 0.4906 &0.3270&0.2534& 0.6993\\
			5 & 0.3845 &0.3927&0.3624&0.2506\\
			6 & 0.1175 &0.2237&0.2786&0.0482\\		
			7 & 0.0067 &0.0521&0.0928&0.0017\\
			8     & 0.0000 &0.0040&0.0126&0.0000\\	
			9&0.0000 &0.0003&0.0000&0.0000\\
	\end{tabular}}
\end{table}

The posterior distribution of number of hidden states for different cases are shown in Table \ref{simex1}, with the trace plots displayed in the Supplement. The trace plots demonstrate that our reversible jump MCMC algorithm has extensively explored the parameter space. The posterior modes for Normal with $\sigma=1$ and Poisson cases are both four, indicating that the proposed MCMC algorithm can identify the number of states where the data are simulated from. However, when we increase $\sigma$ to 1.5 and 2, the posterior modes for the two cases are five. In those cases, the distributions of the number of hidden states are more diverse. In the trace plots for those two cases, the MCMC sampler is more likely to explore the higher dimensional parameter space, resulting in fewer iterations of the four-state model.

\subsection{Replications: Identifying the number of states}
\begin{figure}[ht]
	\centering
	\caption{Posterior distribution of the number of states for Normal case $\sigma=1$ with 100 replications for the same dataset.}
	\includegraphics[scale=0.55]{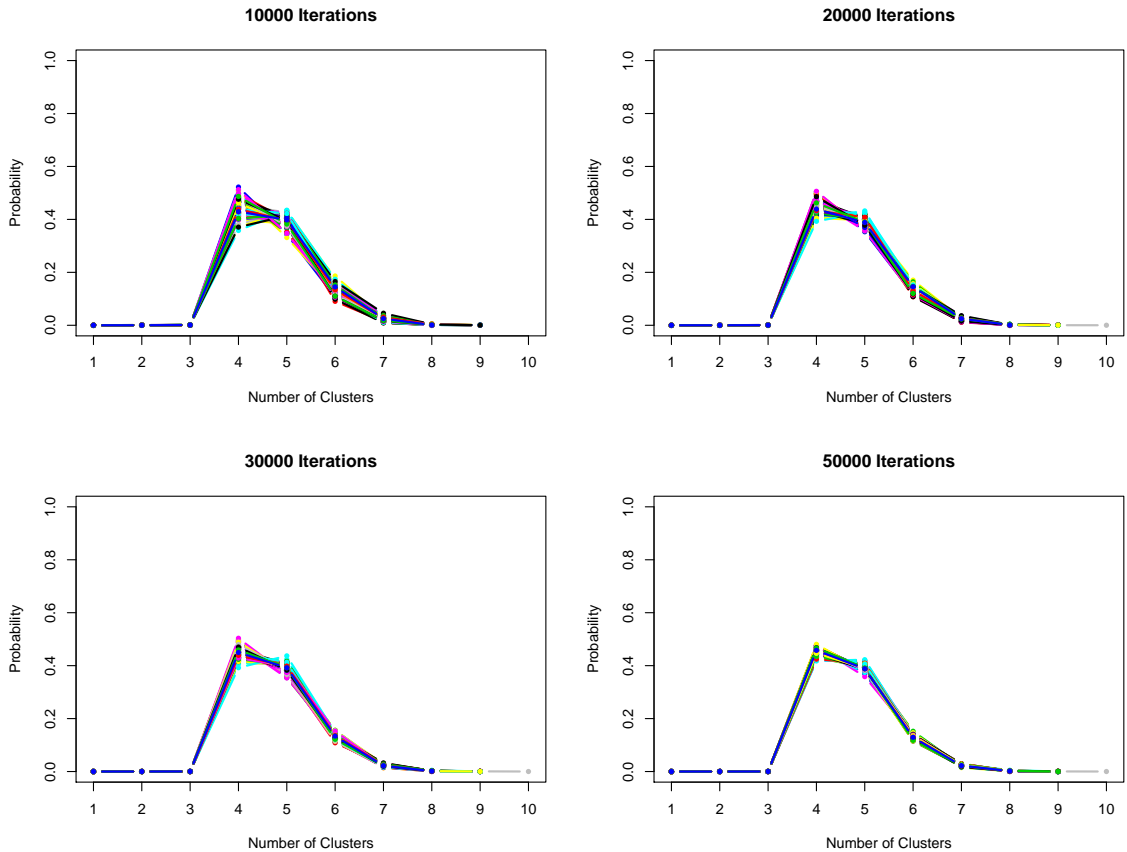}
	\label{Nor1repsamedata}
\end{figure}

Subsequently, we run 100 replications on the same data set with the same parameter configuration and prior settings as Section \ref{sim:rjmcmc} of 500 subjects for Normal case with $\sigma=1$. In each replication, we run 50,000 iterations in total. Figure  \ref{Nor1repsamedata} displays the posterior distribution of the number of states over 100 replications after 10,000, 20,000, 30,000 and 50,000 iterations. In the figure, the proposed RJMCMC algorithm generates consistent results across almost replications, where the majority of them has the posterior mode four. As the number of iterations increases, the variation of the posterior distribution becomes smaller. After 50,000 iterations, 99 out of 100 replications has the posterior mode four, which demonstrates the stability of the proposed algorithm.

\subsection{Identifying the number of states: Intercept Only}
\label{sim:ex2}
In this example, the data are generated with the intercept only in the GLM model. The purpose of this example is to show how the performance  differs from previous examples, especially on the values of $\sigma$ in the Normal case.  The simulation is configured with three latent states and the associated population generator and the coefficient matrix
\begin{equation*}
	Q=\left(\begin{array}{rrr}
		-1.0 & 0.6&  0.4 \\
		0.7 & -1.2  &0.5 \\
		0.3 & 0.6 & -0.9\\
	\end{array}
	\right)
\end{equation*}
with associated coefficient matrices
\begin{itemize}
	\item Gaussian case: $\bB=\left(-4,0,5 \right)$,
	
	\item Poisson case: $\bB=\left(\log(1.5), \log(4), \log(5)\right)$
\end{itemize}
The initial distribution $\pi$ for the continuous-time Markov process is set to be $\left(0.5,0.4,0.1\right)$. As in the first example, we construct the continuous-time Markov process from the generator $Q$, a continuous-time realization of the latent state process $\left\{X_s,0 \le s \le 15\right\}$, and randomly extract observation time points from the $Uniform\left(20,60\right)$ between 0 and 15, with the first observation at time 0 in order to estimate the initial probability $\pi$. We generate data for 1000 subjects in each case. The prior distributions for the elements in $Q$ and $\pi$ are specified as independent $Gamma(1,2)$ and $Dirichlet(1,\ldots,1)$. The priors are imposed for the mean of Normal case as $\mathcal{N}(0,1)$ and  for Poisson case as $Gamma(10,10)$. Again, we use a $Poisson(3.5)$ distribution for as the prior for the number of states, and we initiate the model with one hidden state.

\begin{table}
	\caption{\label{simex2} Example \ref{sim:ex2}: Posterior distribution of the number of states (Intercept Only). The true number of states is three.}
	\centering
	\fbox{%
		\begin{tabular}{*{5}{c}}
			\# of hidden states& Normal $\sigma=1$ & Normal $\sigma=1.5$ & Normal $\sigma=2$ & Poisson\\
			\hline
			1& 0.0001 &0.0001& 0.0001&0.0001\\
			2 & 0.0001&0.0002& 0.0002 &0.0002\\
			3 & 0.9823&0.9671& 0.9533 &0.9533\\
			4 & 0.0174 &0.0328&0.0465& 0.0465\\
			5 & 0.0003 &0.0000&0.0000&0.0000\\
	\end{tabular}}
\end{table}

The results are shown in Table \ref{simex2} with the trace plots of the number of plots in the Supplement. As there are fewer parameters in the example, posterior modes for all cases are three, and all cases have over 90\% of iterations on the three-state model. Unlike previous cases, the algorithm is less likely to explore higher dimensions compared to models with covariates. We notice that the model constantly visit the four-state model then quickly merged back to three-state model, and this prevents the MCMC sampler to move to a higher dimension. In general, the proposed algorithm shows a good mixing performance in this example.

\subsection{Identifying numbers of clusters and states}
\label{mdclustervar}
In this example, we perform a simulation study to examine the performance of the RJMCMC algorithm for clustering trajectories with varying states. We generate the data from a four-cluster CTHMM,  with each cluster having different latent states and specifications
\[
Q_1=\left(\begin{array}{ccc}
-2.5 & 2.0&  0.5\\
0.5 & -1.5  &1.0 \\
0.1&  0.9 & -1  \\
\end{array}
\right)
\;
Q_2=\left(\begin{array}{cc}
-1.20 & 1.20\\
0.25 & -0.25 \\
\end{array}
\right)
\]
\[
Q_3=\left(\begin{array}{ccc}
-0.50 & 0.49&  0.01\\
0.25 & -0.30  &0.05 \\
0.01&  0.10 & -0.11  \\
\end{array}
\right)
\;
Q_4=\left(\begin{array}{cccc}
-3.00 & 2.00&  1.00 & 0.00\\
1.00 & -1.80 & 0.75 & 0.05\\
0.15 & 0.55 &-1.05 & 0.35\\
0.00 & 0.25 & 0.40& -0.65\\
\end{array}
\right)
\]
with associated coefficient matrices
\begin{itemize}[leftmargin=*,noitemsep]
	\small
	\item Gaussian case: $\bB_1=\left(-3,0,2\right)$, $\bB_2=\left(-3.5,3.5\right)$, $\bB_3=\left(-3.8,1,4\right)$, $\bB_4=\left(-2,-1.2,0.7,1.8\right)$.
	
	\item Poisson case: $\bB_1=\left(\log(1.5), \log(4), \log(5)\right)$, $\bB_2=\left(\log(2), \log(6)\right)$, 
	$\bB_3=\left(\log(1.3), \log(4.2), \log(7.5)\right)$, $\bB_4=\left(\log(0.15), \log(0.5),\log(2),\log(6.2)\right)$.
\end{itemize}
The initial distributions for three clusters are $\pi_1=\left(0.5,0.4,0.1\right)$, $\pi_2=\left(0.6,0.4\right)$, $\pi_3=\left(0.45,0.45,0.1\right)$  and $\pi_4=\left(0.35,0.25,0.2,0.2\right)$. We initiate the model with one cluster with one hidden state. Data are generated by constructing the continuous-time Markov chain from the generator $Q_i$ for cluster $i=1,2,3,4$, a continuous-time realization $\left\{X_s,0 \le s \le 15\right\}$, and uniformly extract $T-1$ time points between 0 and 15, where  $T \sim Uniform\left(20,60\right)$.  Data are generated with 400, 500, 450 and 550 subjects for each cluster respectively. We use the same prior distributions with Section \ref{sim:rjmcmc} for the model parameters.

\begin{table}
	\caption{\label{cluvar} Example \ref{mdclustervar}: Posterior distribution of the number of cluster with varying states. The true number of clusters is four.}
	\centering
	\fbox{%
		\begin{tabular}{*{5}{c}}
			\# of cluster& Normal $\sigma=1$ & Normal $\sigma=1.5$ & Normal  $\sigma=2$  & Poisson\\
			\hline
			1& 0.0001 &0.0002& 0.0001& 0.0001\\
			2 & 0.0001&0.0003& 0.0004&  0.0001 \\
			3 & 0.0092&0.0072& 0.0029&  0.0151 \\
			4 & 0.8706 &0.4085& 0.3390&  0.8713\\
			5 & 0.1130 &0.4354&  0.2640& 0.1130\\
			6 & 0.0070&0.1016 & 0.1930& 0.0004\\
			$\ge 7$ &0.0000& 0.0468  &  0.2006 &  0.0000   \\
	\end{tabular}}
\end{table}

Trace plots for the number of clusters for different cases are shown in the Supplement. For the number of clusters, all cases, expect Normal $\sigma=1.5$, have posterior mode four which is the true number of clusters where the data are generated from. For Normal cases, we observe a monotonic decreasing trend for posterior probabilities of four clusters as $\sigma$ increases. The results for the Poisson case are similar to Normal $\sigma=1$. Conditional on four-cluster iterations,  trace plots of the number of states display in the Supplement, with missing parts representing non-four-cluster iterations. The posterior modes of the number of states conditional on four-cluster models are consistent with  where the data are generated from. For Normal $\sigma=1$ and Poisson cases, trace plots of the number of states are similar to the  one-cluster example. For Normal $\sigma=1.5, 2$,   we do not observe, in these two cases, mixing as well as previous examples and there are also fewer four-cluster iterations. When $\sigma=1.5$, the posterior modes of the number of states conditional on the four-cluster model are still the consistent with the true data configuration; however, when $\sigma=2$, it is not easy to identify the number of states in each cluster. Compared to previous cases, this is a more difficult problem because of the complexity and the flexibility of the proposed algorithm. For example, when updating the number of clusters, it is less likely to have a successful combine move until two similar clusters have the same number of states. In our example, we set the probability of the combine move for updating the number of clusters as 0.7 to account for issue. Overall, this algorithm performed well in selecting the number of clusters and states in well-separated scenarios.

\section{Real Data Analysis: Health Surveillance of COPD patients}
\label{real}
Our real example relates to healthcare surveillance for the chronic condition, COPD, in greater Montreal area, Canada.  In 1998, a 25\% random sample was drawn from the registry of the R\'{e}gie de l'assurance maladie du Qu\'{e}bec (RAMQ, the Qu\'{e}bec provincial health authority) with a residential postal code in the census metropolitan area of Montreal. At the start of every following year, 25\% of those who were born in, or moved to, Montreal within the previous year were sampled to maintain a representative cohort. Follow-up ended when people died or changed their residential address to outside of Montreal. This administrative database includes outpatient diagnoses and procedures submitted through RAMQ billing claims, and procedures and diagnoses from inpatient claims.

Using established case-definitions based on diagnostic codes \citep{lix2018canadian},  COPD  patients were enrolled with an incident event occurring after a minimum of two years at risk with no events. Patients were followed from January 1998, starting from the time of their first diagnosis, until December 2014.  Physicians only observed these patients during medical visits, which occurred when patients chose to interact with the healthcare system, and at which information, including the number of prescribed medications, is collected.  However, as this information was only available for patients with drug insurance, we restrict the cohort to patients over 65 years old with COPD, as prescription data are available for all of these patients. It is widely believed that the progression of COPD can be well-modeled as a progression through a small number of discrete states which approximate severity \citep{r12}. We are interested in identifying those states and modeling transition between these discrete states, which reflects the performance of the healthcare system over time.

In our analysis, the outcome observations are the number of prescribed medications at the time when patients visited the physician: these are modeled using a Poisson model. In addition, the types of healthcare utilization at each visit were also recorded: hospitalization (HOSP), specialist visit (SPEC), general practitioner visit (GP) and emergency department visit (ER). 4,597 COPD patients are included in this analysis, and these patients are all with drug plans and with at least five years follow-up.

\subsection{Identifying the number of states}
\label{copdstat}
First, we carry out our analysis to identify the number of states. The analysis is initiated as a one-state model,  The prior distributions for the elements in $Q$ and $\pi$ are specified as independent $Gamma(1,2)$ and $Dirichlet(1,\ldots,1)$. We use a $Poisson(3)$ distribution for as the prior for the number of states.
\subsubsection{With covariates}
\label{withcov}
We implement the model including the types of healthcare utilization as covariates in the observation model. A non-informative prior is imposed for $\bB$. We perform the proposed trans-dimensional MCMC algorithm with 20,000 iterations.
\begin{table}
	\caption{\label{realnum}Application: Posterior distribution of the number of states corresponding to models with and without healthcare utilizations as a covariate in   Sections \ref{withcov} and  \ref{inter}, respectively. }	
	\centering
	\fbox{%
		\begin{tabular*}{37.5pc}{@{\hskip5pt}@{\extracolsep{\fill}}c@{}c@{}c@{}c@{}c@{}c@{}c@{}c@{}c@{\hskip5pt}}
			\# of states& 1 & 2 & 3 & 4 & 5 & 6&7&8\\
			\hline
			With Covariates& 0.0001 &0.0002& 0.0002&0.4212&\textbf{0.4526} &0.0796 &0.0415& 0.0045 \\
			Without Covariates& 0.0001 &0.0005& 0.3863&\textbf{0.4808}&0.1261 &0.0064 &0.0000& 0.0000 \\
	\end{tabular*}}
\end{table}

\begin{figure}[ht]
	\centering
	\caption{Application: Trace plot for the number of states over 20000 iterations to identify the number of states. The left panel is the observation model with the types of healthcare utilization as covariates, while the right panel is the model without covariates.}
	\begin{minipage}[b]{0.4\textwidth}
		\includegraphics[width=\textwidth]{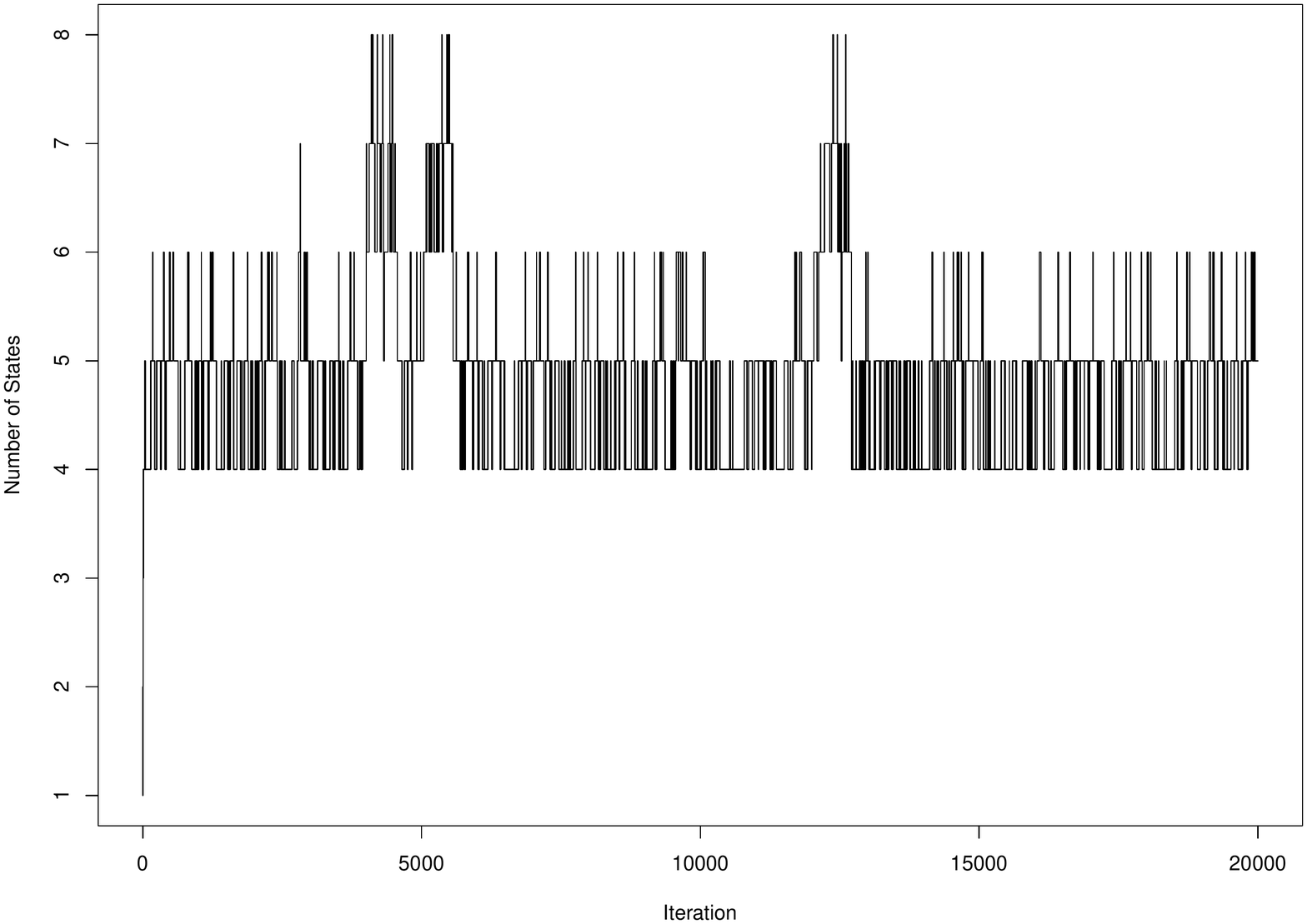}
	\end{minipage}
	\hfill
	\begin{minipage}[b]{0.4\textwidth}
		\includegraphics[width=\textwidth]{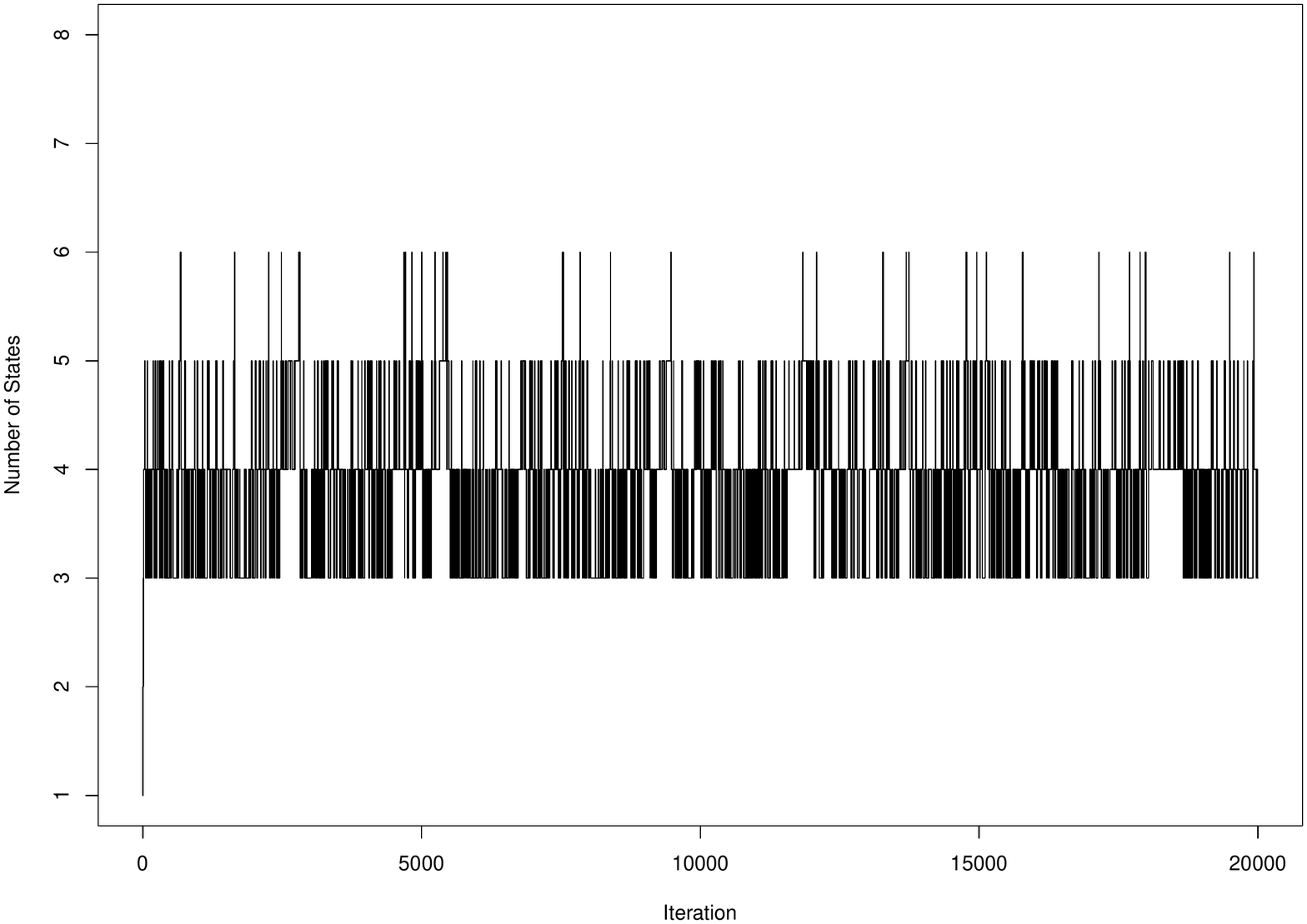}
	\end{minipage}
	\label{tracecopd}
\end{figure}

\begin{table}
	\caption{\label{realdrug}Application: Exponential of $\bB$ coefficients (Parameters in the GLM for each state).}
	\centering
	\fbox{%
		\begin{tabular}{*{6}{c}}
			Variables & State 1 &  State 2 &  State 3 & State 4 & State 5 \\
			\hline
			Intercept &2.05 & 3.53 & 4.55 & 6.03& 7.52\\
			(95\% CI) & (1.80,2.34)  & (3.00,4.02) & (3.74,5.32) & (5.35,6.87) & (7.02,8.14) \\
			ED  &  1.03 & 1.00 & 0.99 & 0.98& 0.97  \\
			(95\% CI) & (1.00,1.06) & (0.98,1.02)& (0.97,1.00)& (0.97,0.99)&(0.95,0.99)\\
			HOSP &  1.00 & 1.00 & 0.99 &0.98 &0.96 \\
			(95\% CI) & (0.95,1.05) & (0.95,1.02)& (0.95,1.01) & (0.97,1.00)&(0.92,0.99) \\		
			SPEC &1.02 &0.98&0.96&0.97&0.93\\
			(95\% CI) & (0.98,1.06) & (0.96,1.01) & (0.94,0.98)&(0.95,0.98)&(0.89,0.96)\\
	\end{tabular}}
\end{table}

Table \ref{realnum} shows the posterior distribution of the number of states. The trace plot (Figure \ref{tracecopd}) confirm that  the algorithm has fully explored the parameter space.  Although the mode of the posterior distribution of the number of states is five, it also spends over 40\% of iterations in the four-state model.  Table \ref{realdrug} contains the exponential of the $\bB$ coefficients condition on the five-state model. On average, from State 1 to 5 the number of drugs taken increases; however, within each state, the numbers of drugs across the different healthcare utilizations are approximately the same. Therefore, it is plausible to consider  fitting the intercept-only model without the time-varying covariate, which we will proceed in the next section. 

\subsubsection{Without covariates}
\label{inter}
We perform the reversible jump trans-dimensional MCMC algorithm for 20,000 iterations without the time-varying covariate, with a $Gamma(10,10)$ distribution placed on the mean number of drugs.  Table \ref{realnum} shows the posterior distribution of the number of states. The trace plot (Figure \ref{tracecopd}) confirm that  the algorithm has extensively explored the parameter space. Unlike the model with  the time-varying covariate, the MCMC algorithm employs most of the time exploring the less complex models, i.e., three-state and four-state model. The posterior mode of the number of hidden states is four. Table \ref{realdrug2} contains the expected number of drugs prescribed for patients in each state, with associated 95\% credible intervals.  As for the model with covariates included, on average,  the number of drugs taken increases from State 1 to 4; however, the mean number of drugs prescribed for each state is smaller than the previous five-state model with the time-varying covariate.

\begin{table}
	\caption{\label{realdrug2}Application: Expected number of drugs for the intercept-only model over the time spent in each state.}
	\centering
	\fbox{%
		\begin{tabular}{*{5}{c}}
			& State 1 &  State 2 &  State 3 & State 4 \\
			\hline
			Expected \# of Drug Prescribed&3.19 & 4.00 & 4.75 & 5.90\\
			(95\% CI) & (2.89,3.31)  & (3.29,4.58) & (4.53,5.84) & (5.85,6.05) \\
	\end{tabular}}
\end{table}

\subsection{Identifying numbers of clusters and states}
Next, we implement the clustering algorithm to group trajectories with distinct stochastic properties.  From the previous one-cluster model, we did not observe much distinction across different healthcare utilizations on the number of drugs. Therefore, we decide to cluster patient trajectories using the intercept-only model.

\begin{table}
	\caption{\label{cluvarcopd} Application: Posterior distribution of the number of cluster and numbers of states conditional on three-cluster iterations}
	\centering
	\fbox{%
		\begin{tabular}{*{5}{c}}
			&Number of clusters& \multicolumn{3}{c}{Number of states}\\
			&  & Cluster 1 & Cluster 2  & Cluster 3 \\
			\hline
			1& 0.0058 &0.0000& 0.0000& 0.0000\\
			2 & 0.3678&0.0000& \textbf{0.7413}&  \textbf{0.6480} \\
			3 &\textbf{0.5358} &0.0418& 0.2447&  0.1566 \\
			4 & 0.0820  &\textbf{0.9580}& 0.0140&  0.1627\\
			5 & 0.0072 &0.0002&  0.0000& 0.0269\\
			6 & 0.0014&0.0000 & 0.0000& 0.0058\\
	\end{tabular}}
\end{table}
\begin{figure}[ht]
	\centering
	\caption{Application: Trace Plot for the Number of Clusters}
	\includegraphics[scale=0.55]{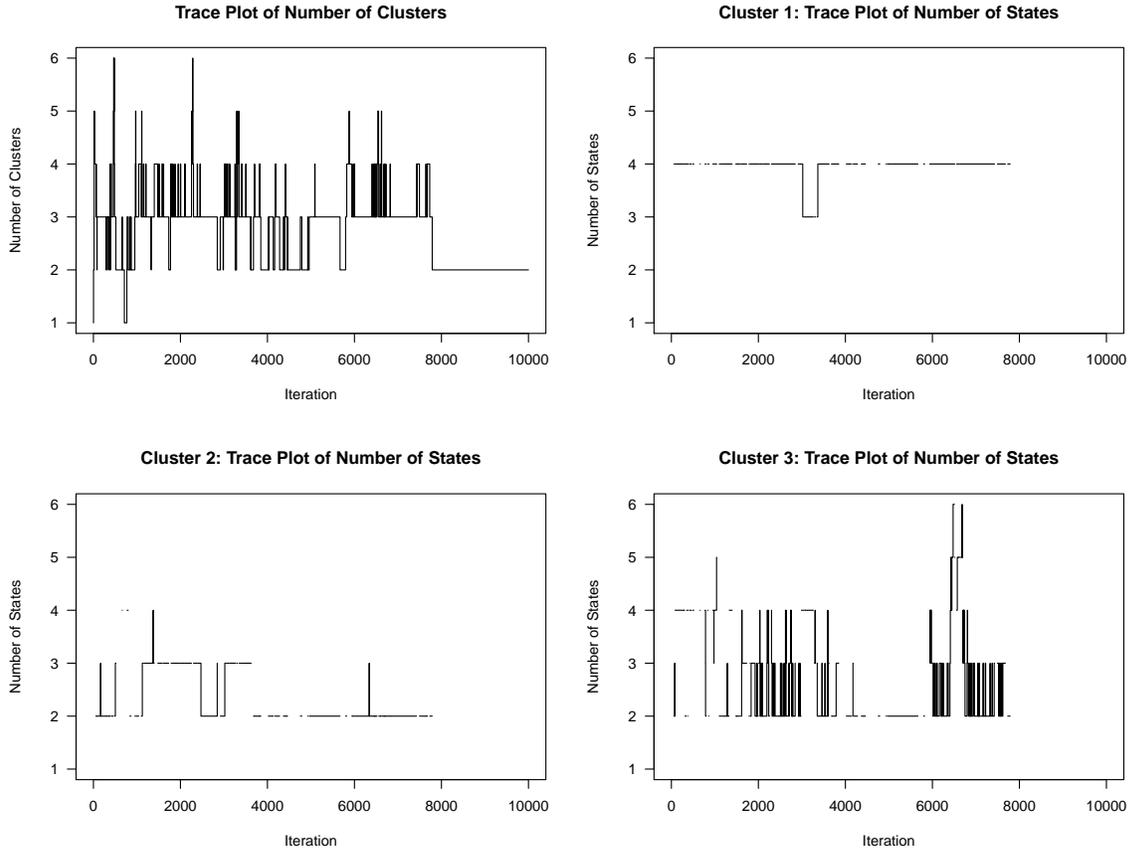}
	\label{tracecopdcl}
\end{figure}

We present results based on 10000 MCMC iterations after initialization from one-cluster model with one hidden state. The mode of the posterior distribution of the number of clusters is three (5358 out of 10000 iterations). Table \ref{cluvarcopd} and Figure \ref{tracecopdcl} present the posterior distribution and trace plots of the number of clusters and numbers of states conditional on three-cluster iterations. The posterior modes for numbers of states are four, two and two for Cluster 1, 2, 3 respectively. For a summary output, cluster membership is assigned to the subject according to its posterior mode conditional on three-cluster iterations. Table \ref{copdcl} shows the posterior mean of number of drugs for the three-cluster model along with the number of patients in each cluster. Cluster 1 has the greatest number of patients and a posterior mode  of four states, which is consistent with results of the one-cluster model in Section \ref{copdstat}. The separation between Cluster 2 and 3 is mainly coming from the parameters in the underlying Markov process, as the $q_{12}$ and $q_{21}$  in Cluster 3 are ten times greater than those in Cluster 2.  This suggests that transitions between State 1 and 2 are more frequent in Cluster 3. Also, Cluster 3 on average has the least number of drugs prescribed, indicating that patients in this cluster are possibly on the early stage of COPD.

\begin{table}
	\caption{\label{copdcl}Application: Expected number of drugs for the three-cluster Poisson model.}
	\centering
	\fbox{%
		\begin{tabular*}{37.5pc}{@{\hskip5pt}@{\extracolsep{\fill}}c@{}c@{}c@{}c@{}c@{}c@{\hskip5pt}}
			&	& State 1 &  State 2 &  State 3 & State 4 \\
			\hline
			Cluster 1 	& \# of Drug Prescribed&2.04&  3.38& 4.79&6.43\\
			$N=4439$	&		(95\% CI) & (1.93,2.26)  & (3.04,3.77) & (4.42,5.28) & (6.12,6.79) \\
			\hline
			Cluster 2	& \# of Drug Prescribed&3.72& 6.62  & &\\
			$N=135$	&		(95\% CI) & (3.13,4.48)  & (4.19,7.74) &  &  \\
			\hline 
			Cluster 3	&\# of Drug Prescribed&3.15&  5.61& &\\
			$N=23$	&		(95\% CI) & (2.48,5.47)  & (4.09,7.66) & & \\
		\end{tabular*}
	}
\end{table}
\section{Discussion}
\label{dis}
We have developed a reversible jump MCMC algorithm for the CTHMM-GLM with an unknown number of states and clusters, which is implemented under a fully Bayesian framework.  This model can deal with challenges typically encountered in latent multi-state modeling, in particular, irregular visits that vary from individual to individual. Our approach uses a split/combine move to explore the trans-dimensional parameter space, which extended the fixed dimensional MCMC proposed by \cite{luocthmm2018a}.  Simulation studies demonstrated that the proposed MCMC approach could identify the number of states and the number of  clusters from the true data generating mechanism.  We were able to implement the developed methods for a real data set from Quebec, Canada, comprising more than four thousand COPD patients tracked over twenty years. Our work demonstrated that with a careful construction of the trans-dimensional proposal, our reversible jump MCMC algorithm can achieve desired performance in term of identifying the number of states and the number of clusters simultaneously.  

Focusing on the number of states and the number of clusters, a standard prior specification is adopted  exchangeable in form with respect to the state/cluster labels. In the MCMC algorithm, it is possible that the algorithm would potentially suffer from the label-switching problem, which has been addressed by \cite{jasra2005markov}. In this paper, we primarily considered  finite mixture formulations to  facilitate the trans-dimensional move between different numbers of states and clusters using the reversible jump MCMC algorithm.  Bayesian nonparametric procedures, specifically procedures using Dirichlet process models, have become popular tools to explore the trans-dimensional parameter space, where the models are limiting versions of exchangeable finite mixture models. Dirichlet process models are now widely used in density estimation and clustering, with implementation via MCMC sampling approaches \citep{neal2000markov}.  It would be interesting to apply Dirichlet process models for CTHMMs to identify the number of states and the number of clusters simultaneously  and compare aspects of finite and Dirichlet process mixture formulations, noting their similarities and where they differ. In our proposed model, the state space has to be discrete; more generally, there may be health conditions that necessitate the use of a continuous latent process. Bayesian formulations for diffusion or jump processes have been studied in the context of financial data, although such formulations are not common in the analysis of health data, allowing the latent continuous state distribution to have an interpretation as an index or a score. For example, one could use the features included in comorbidity indices to measure multimorbidity in terms of the ability to predict future mortality and health services use. Further studies are needed to address this issue to  facilitate the generation of hypotheses about the performance of the healthcare system in managing patients with chronic disease.

\bibliographystyle{chicago}
\bibliography{CTHMMunknown_bib.bib}

\clearpage

\appendix

\begin{center}
	{\LARGE\bf Supplementary Materials for ``Bayesian inference for continuous-time hidden Markov models with an unknown number of states"}
\end{center}

\section{Likelihood for a Continuous-Time Hidden Markov Model}

Figure \ref{tikzfig:data} provides a schematic of the presumed data generating structure for one subject.

\tikzset{filled/.style={fill=circle area, draw=circle edge, thick},
	outline/.style={draw=circle edge, thick}}

\tikzset{cross/.style={cross out, draw=black, minimum size=2*(#1-\pgflinewidth), inner sep=0pt, outer sep=0pt},
	cross/.default={1pt}}
\begin{figure}[ht]
	\centering
	\begin{tikzpicture}
	
	\draw[decorate,
	decoration={zigzag,
		pre length=5.5cm,
		post length=2cm,
		amplitude=2mm
	}] (0,0) -- (10,0);
	\draw[decorate,
	decoration={zigzag,
		pre length=5.5cm,
		post length=2cm,
		amplitude=2mm
	}] (0,1) -- (10,1);
	\draw[decorate,
	decoration={zigzag,
		pre length=5.5cm,
		post length=2cm,
		amplitude=2mm
	}] (0,2) -- (10,2);
	
	\draw (0,0)  node[cross=3pt]{};
	\draw (0,0.1)  node[above] {1};
	\draw (2,0)  node[cross=3pt]{};
	\draw (2,0.1)  node[above] {2};
	\draw (2.5,0)  node[cross=3pt]{};
	\draw (2.5,0.1)  node[above] {3};
	\draw (4.1,0)  node[cross=3pt]{};
	\draw (4.1,0.1)  node[above] {2};
	\draw (9,0)  node[cross=3pt]{};
	\draw (9,0.1)  node[above] {1};

	\fill (0,2)  circle[radius=2pt] node[above] {$\tau_1=0$};
	\fill (1,2)  circle[radius=2pt] node[above] {$\tau_2$};
	\fill (1.5,2)  circle[radius=2pt] node[above] {$\tau_3$};
	\fill (2.4,2)  circle[radius=2pt] node[above] {$\tau_4$};
	\fill (4.5,2)  circle[radius=2pt] node[above] {$\tau_5$};
	\fill (8.5,2)  circle[radius=2pt] node[above] {$\tau_T$};
	
	\draw (0,1)  circle[radius=2pt] node[above] {$O_1$};
	\draw (1,1)  circle[radius=2pt] node[above] {$O_2$};
	\draw (1.5,1)  circle[radius=2pt] node[above] {$O_3$};
	\draw (2.4,1)  circle[radius=2pt] node[above] {$O_4$};
	\draw (4.5,1)  circle[radius=2pt] node[above] {$O_5$};
	\draw (8.5,1)  circle[radius=2pt] node[above] {$O_T$};
	
	\node[text width=1cm] at (-0.5,2) {$s$};
	\node[text width=1cm] at (-0.5,1) {$O_s$};
	\node[text width=1cm] at (-0.5,0) {$X_s$};
	
	\end{tikzpicture}
	\caption{Schematic of the presumed data generating mechanism for the CTHMM.  $O_s$ represents the process underlying the observed data, with observation time points denoted $\tau_t, t=1,\ldots,T$; $X_s$ represents the hidden Markov process.\label{tikzfig:data}. Reproduced from \cite{luo2019bayesian}.}
\end{figure}

To construct the complete data likelihood, suppose that $\{X_s\}$ has been observed continuously in the time interval $\left[0,\tau \right]$.  The likelihood function for $Q$ for individual $n$ is \citep{bladt2005statistical}
\begin{equation*}
	\label{difflik}
	\mathcal{L}_{n,l,m}(\tau) = \prod\limits_{l = 1}^K {\prod\limits_{m \ne l} {{q_{l,m}}^{{N_{n,l,m}}\left( \tau  \right)}\exp \left( { - {q_{l,m}}{R_{n,l}}\left( \tau  \right)} \right)} }
\end{equation*}
where $N_{n,l,m} ( \tau )$ is the number of transitions from state $l$ to state $m$ in the time interval $\left[0,\tau\right]$ and $R_{n,l}\left( {{\tau}} \right)$ is the total time that the process has spent in state $l$ in $\left[0,\tau\right]$ for individual $n$, and where $q_{l,m}$ are the potentially elements of $Q$.   Note that the quantities $N_{n,l,m} \left( \tau  \right)$ and $R_{n,l} \left( \tau  \right)$ are unobserved, but can be computed given a complete realization of the latent process on $[0,\tau]$.

To construct the likelihood for $\Theta$ for $N$ independent subjects, we let $O_{n,t}$ $\left(t=1,\ldots,T_n\right)$  be the $t^{\text{th}}$ observation for subject $n$ with the associated observation time $\tau_{n,t}$. The complete data likelihood derived from $\left\{O_{n}\right\}$ and $\left\{X_{n,\tau_n}\right\}$ can be factorized $\mathcal{L}(\Theta) \equiv \mathcal{L} (\bO, \bX | \Theta ) =  \mathcal{L} (\bX | \Theta ) \mathcal{L} (\bO | \bX, \Theta )$ where $\bO = \{O_{n,t}\}$ and $\bX = \{X_{n,\tau_{n,t}}\}$ for $n=1,\ldots,N$ and $t=1,\ldots,T_n$ and
\begin{equation*}
	\begin{aligned}
		&\mathcal{L} (\bO  |  \bX, \Theta )
		=\prod\limits_{n=1}^N \prod\limits_{t = 1}^{T_n} {f\left( {{O_{n,t}}\left| {{X_{n,\tau_{n,t}}}} \right.} \right)}\\
		&\mathcal{L} (\bX | \Theta )  = \prod\limits_{n=1}^N {\pi_{X_{n,0}}} \left\{ \prod\limits_{t = 1}^{{T_n}  - 1} \mathcal{L}_{n,l,m}(\Delta_{n,t}) \right\}
	\end{aligned}
\end{equation*}
where $\Delta _{n,t} = \tau_{n,t+1}-\tau_{n,t}$.  The complete data log-likelihood written in terms of the latent state indicator random vectors $\left\{S_k\right\}_{k=1}^{K}$ is
\begin{align*} \label{loglik}
	\ell (\Theta ) & =\sum\limits_{n=1}^N {\sum\limits_{t = 1}^{T_n}  {\sum \limits_{k=1}^K{S_{n,t,k} \log f\left( {{O_{n,t}}\left| {{S_{n,t,k}}} \right.} \right)}}} + \sum\limits_{n=1}^N { \sum \limits_{k=1}^K S_{n,1,k} \log\left( \pi_{k}\right)} \\
	& \qquad \qquad \qquad + \sum\limits_{n=1}^N \sum\limits_{t = 1}^{{T_n} -1}{\sum \limits_{j=1}^K  \sum \limits_{k=1}^K S_{n,t,j}S_{n,{t+1},k}  p_{n,t}^{j,k} } \nonumber
\end{align*}
where
\[
p_{n,t}^{j,k} = \sum\limits_{l = 1}^K \sum\limits_{m \ne l} \left\{N_{n,l,m}^{j,k}\left( \Delta _{n,t} \right) \log \left( q_{l,m} \right)- q_{l,m} R_{n,l}^{j,k} \left( \Delta _{n,t} \right) \right\}.
\]
records the probability of transition from state $j$ to state $k$ in the interval $\Delta _{n,t}$, and $N_{n,l,m}^{j,k}\left( \Delta _{n,t} \right)$ and $R_{n,l}^{j,k} \left( \Delta _{n,t} \right)$ are the amended versions of $N_{l,m}$ and $R_l$ computed conditional on starting in state $j$ and ending in state $k$ over the interval $\Delta _{n,t}$.

Bayesian inference for this model with the number of states $K$ fixed has been fully studied by \cite{luocthmm2018a}, where an MCMC scheme based on simulating the complete latent path for each individual is developed; this MCMC scheme relies upon the rejection sampling approach of \cite{hobolth2009simulation} to sample the latent paths in an efficient fashion. Bayesian inference using the complete data likelihood formulation is appealing as it produces posterior samples of the full unobserved state sequences and latent continuous time process, which allows inferences to be made for individual-level trajectories across the entire observation window, and which is useful for computing posterior distributions for pathwise aggregate features on the individual trajectories.

\section{Updating Model Parameters for CTHMMs with a Fixed Number of States}
\label{sec:fixpara}
For fixed $K$, we may use a standard a Metropolis-Hastings-within-Gibbs algorithm to generate samples from the posterior distribution for $\Theta_K$. Starting with initial values $\pi^{\left(0\right)}$, $\left\{q_{i,j}\right\}_{1\le i\ne j\le K}^{\left(0\right)}$ and $B^{\left(0\right)}$, and then given those initial values, the `forward' and `backward' values
\[
a_{n,t,k} = \mathbb{E} [S_{n,t,k} |\bo;\Theta_K ] = \mathbb{P} (S_{n,t,k}=1 | \bo;\Theta_K) = \sum\limits_{j=1}^K { b_{n,t,k,j}^{\text{old}}}
\]
and $b_{n,t,k,j} = \mathbb{P} (S_{n,t,k}=S_{n,{t+1},j}=1 |\bo;\Theta_K )$ are calculated using the forward-backward algorithm \citep{r5,r7}; see \cite{luocthmm2018a} for specific details.  At iteration $i$, the MCMC algorithm simulates the posterior sample based on the full conditional posterior distributions:

\begin{itemize}
	\item \textbf{Update latent state indicators:} For each $n$ and $t$, generate the random vector $S_{n,t}^{\left(i\right)}$ from the multinomial distribution with parameters $a_{n,t}^{\left(i\right)}=(a_{n,t,1}^{\left(i\right)},\ldots,a_{n,t,K}^{\left(i\right)})$.
	
	\item \textbf{Update $\bB$:} Sample coefficient matrix $\bB^{\left(i\right)}$ and scale parameter $\phi^{\left(i\right)}$ given $S_{n,t}^{\left(i\right)}$ via the Metropolis-Hastings algorithm as there is no standard distributional forms for the conditional posteriors of $\bB$ and $\phi$.  Proposals are made using a standard Metropolis update from a Normal density for elements of $\bB$.  Starting values are obtained using an initial GLM fit to the observed data.
	
	\item \textbf{Update $\pi$:} For a  conjugate $Dirichlet\left(\alpha_1,\ldots,\alpha_K\right)$ prior, sample $\pi^{\left(i\right)}$ from a Dirichlet distribution with parameters
	\[
	\left(\sum \limits_{n=1}^{N}{S_{n,1,1}^{\left(i\right)}}+\alpha_1,\ldots,\sum \limits_{n=1}^{N}{S_{n,1,K}^{\left(i\right)}}+\alpha_K\right)
	\]
	
	\item \textbf{Update $Q$:} This update is achieved by augmenting the sample space by simulating a path for the latent process.  For each $n$ and $t$,
	
	\begin{itemize}
		
		\item Sample the current state and next state ($X_{n,\tau_{n,t}}, X_{n,\tau_{n,t+1}}$) from a multinomial distribution with the parameter matrix containing the $b_{n,t,k,j}$. Since the likelihood for $Q$ requires continuously observed Markov chain, we first simulate the full path before updating $Q$.
		
		
		\item Simulate ${N_{n,l,m}}\left( {{\Delta _{n,t}}}\right)$ and ${R_{n,l}}\left( {{\Delta _{n,t}}}\right)$ from the Markov jump processes step-by-step with infinitesimal generator $Q^{\left(i-1\right)}$ through the intervals $\left[\tau_{n,t},\tau_{n,t+1}\right)$ initiated at $X_{n,\tau_{n,t}}$ and end point $X_{n,\tau_{n,t+1}}$ sampled previously. 	Simulating sample paths conditional on the endpoints can be achieved efficiently by using modified rejection sampling (that avoids simulating constant sample paths when it is known that at least one state change must take place) as proposed by \cite{hobolth2009simulation}, from which $\left\{X_s\right\}$ is recovered and the jump time points are generated.
		

		\item Sample the $\left\{q_{i,j}\right\}_{1\le i\ne j\le K}^{\left(i\right)}$ given the fully recovered Markov process from independent Gamma distributions with shape and rate parameters given as
		\[
		\textrm{shape} = \sum\limits_{n=1}^N {\sum\limits_{t = 1}^{T_n} { {N_{n,l,m}}\left( {{\Delta _{n,t}}} \right)} }+1 \qquad \textrm{rate} = \sum\limits_{n=1}^N {\sum\limits_{t = 1}^{T_n} { {R_{n,l}}\left( {{\Delta _{n,t}}} \right)} }.
		\]
		
	\end{itemize}
	
\end{itemize}

\section{Birth-Death MCMC for CTHMMs}
\label{dbmcmc}
In this section, an alternative approach to infer the number of hidden states via a birth-death process is introduced. Instead of constructing a reversible jump approach, \cite{stephens2000bayesian} described a birth-death method, which view each component of the mixture as a point in the parameter space. The birth or death of a state occurs as a marked point process. The detailed description of one iteration of the MCMC algorithm to incorporate the birth-death process is as  follows:

\begin{enumerate}
	\item Given the state of the parameter $\Theta^{(t)}$ at time $t$, sample the $\Theta^{(t)'}$ by running the birth-death process for a fixed time $t_0$. Set $K^{(t+1)}=K^{(t)'}$.
	\item Fix the number of states $K$. Update the following parameters given $\Theta^{(t)'}$.
	\begin{itemize}
		\item Update latent state indicators $S_{n,t}$.
		\item Update the parameters associated with the observation process $\bB$.
		\item Update the initial distribution $\pi$.
		\item Update the infinitesimal generator $Q$.
	\end{itemize}
\end{enumerate}
As discussed in \cite{stephens2000bayesian}, with a fixed $K$, Step 2 helps improve the mixing of MCMC.
\subsection{Birth-Death Process}
The birth or death of a component occurs as a Poisson process. Let the birth rate as $\lambda_b$. According to Theorem 3.1 in \cite{stephens2000bayesian}, to ensure that a Markov jump process has an invariant probability density which is proportional to the posterior distribution, it is sufficient that
\begin{align*}
	&p_0\left(K\right)p_0\left(\Theta_{K}\right)\mathcal{L}\left(K,\Theta_K\right)\q\left((K,\Theta_K) \to(K+1,\Theta_{K+1}) \right)\\
	&=p_0\left(K+1\right)p_0\left(\Theta_{K+1}\right)\mathcal{L}\left(K+1,\Theta_{K+1}\right)\q\left((K+1,\Theta_{K+1}) \to(K,\Theta_{K})
	\right)
\end{align*}
where $q\left((K+1,\Theta_{K+1}) \to(K,\Theta_{K})\right)$ is the proposal density from $(K+1,\Theta_{K+1}) \to(K,\Theta_{K})$.
Hence, the corresponding death rate for State $K$, given $K+1$ states, should be
\[
\delta_j=\lambda_b\frac{p_{0Q}\left(q'_{K,K+1}\right)p_{0Q}\left(q'_{K+1,K}\right)\prod\limits_{i=1}^{K-1}p\left(w_i\right) p\left(w\right) }{\prod\limits_{i=1}^{K-1}q_{i,K} \times \pi_{K}}\times  r\left( K,\Theta_{K};K+1,\Theta_{K+1}|\bo \right)
\]
The total death rate is $\delta_d=\sum_{j=1}^{K+1}\delta_j$.
Once a birth or death happens, the time to next birth or death follows an exponential distribution with rate $\delta_d+\lambda_b$, with respective probabilities
\[
\mathbb{P}\left(\text{birth move}\right)=\frac{\lambda_b}{\delta_d+\lambda_b} \quad \quad \quad \mathbb{P}\left(\text{death move}\right)=\frac{\delta_d}{\delta_d+\lambda_b}
\]
\begin{itemize}
	\item If the birth move is simulated, sample the new component parameters according to the split move in reversible jump MCMC.
	
	\item If the death move is simulated, select a hidden state to `die' with probability $\delta_j/\delta_d$ where $j=1,\ldots,K+1$.
	\item Simulate the time to the next jump from an exponential distribution with rate $\delta_d+\lambda_b$.
\end{itemize}

\section{Simulation: Birth-Death MCMC}
We used the exactly the same data generating mechanism in Section 5.3 in the main paper. The prior distributions are also the same with that example. We initiate the model with one hidden state. We update the number of hidden states using the birth-death process with a fixed time $t_0$, with a birth rate $\lambda_b=1$.

\begin{table}
	\caption{\label{simex3} BDMCMC: Posterior distribution of the number of states (Intercept Only)}
	\centering
	\fbox{%
		\begin{tabular*}{37.5pc}{@{\hskip5pt}@{\extracolsep{\fill}}c@{}c@{}c@{}c@{}c@{}c@{\hskip5pt}}
			$t_0$	&\# of hidden states& Normal $\sigma=1$ & Normal $\sigma=1.5$ & Normal $\sigma=2$ & Poisson \\
			\hline
			\multirow{7}{*}{$10^{-15}$}&	1& 0.0001 &0.0001& 0.0001&0.0001 \\
			&	2 & 0.0001&0.0001& 0.0001 &0.0002\\
			&	3 & 0.4508&0.2096& 0.3313 &0.2967\\
			&	4 & 0.4932 &0.4529&0.4758& 0.4704\\
			&		5 & 0.0506 &0.2900&0.1762&0.2023\\
			&		6 & 0.0054 &0.0470&0.0166&0.0298\\
			&		7 & 0.0000 &0.0004&0.0001&0.0008\\
			\hline
			\multirow{7}{*}{$10^{-20}$}&	1& 0.0001 &0.0001& 0.0001&0.0001 \\
			&	2 & 0.0001&0.0001& 0.0001 &0.0001\\
			&	3 & 0.4529&0.3899& 0.3334 &0.4083\\
			&	4 & 0.4976 &0.4881&0.4798& 0.4904\\
			&		5 & 0.0481 &0.1122&0.1677&0.0920\\
			&		6 & 0.0014 &0.0096&0.0191&0.0091\\
			&		7 & 0.0000 &0.0001&0.0000&0.0000\\	
	\end{tabular*}}
\end{table}

\begin{figure}[ht]
	\centering
	\caption{Trace Plots for the Number of States of 20000 Iterations using BDMCMC with $t_0=10^{-15}$ (left panel) and $t_0=10^{-20}$ (right panel).}
	\begin{minipage}[b]{0.45\textwidth}
		\includegraphics[width=\textwidth]{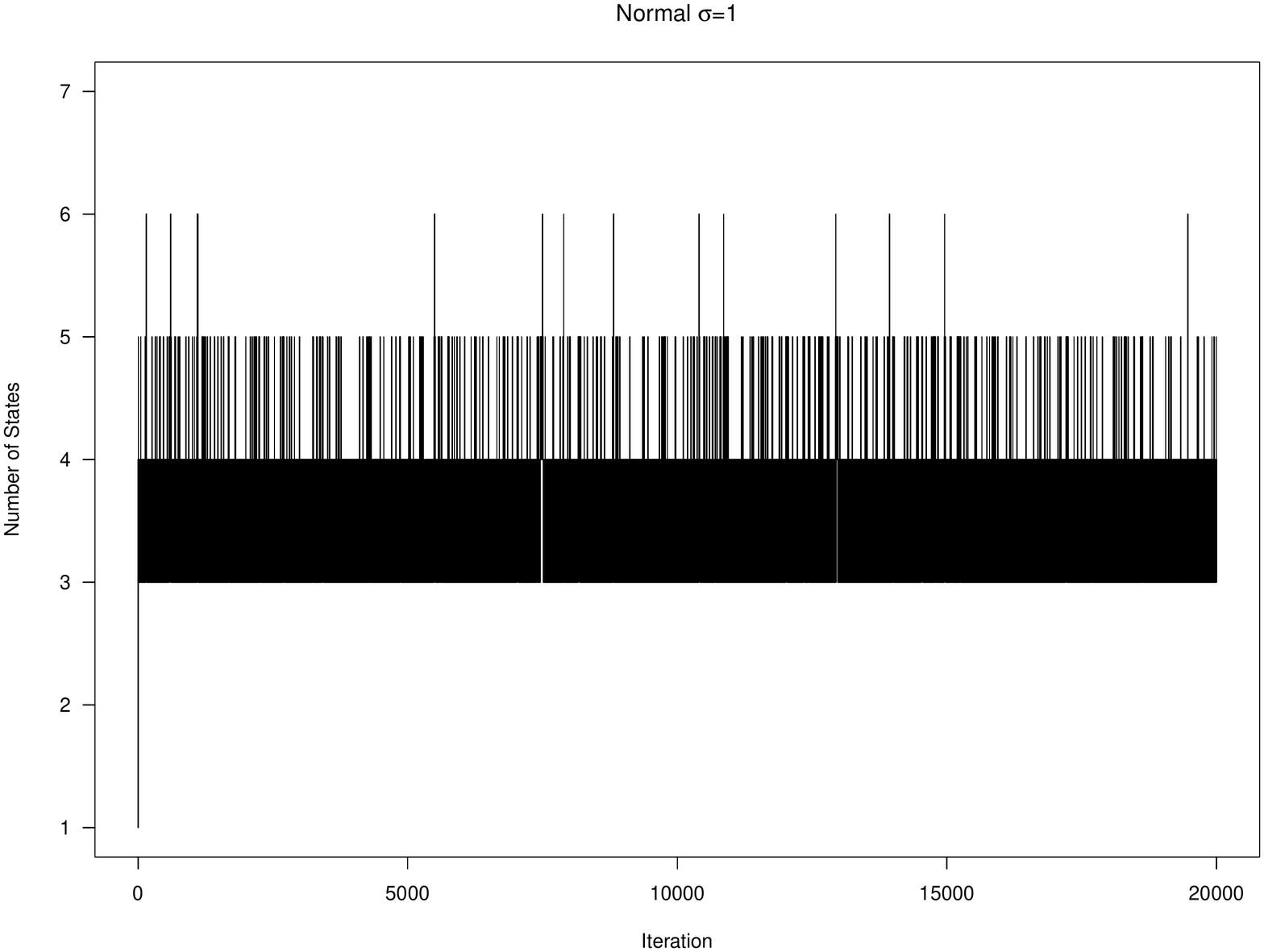}
	\end{minipage}
	\hfill
	\begin{minipage}[b]{0.45\textwidth}
		\includegraphics[width=\textwidth]{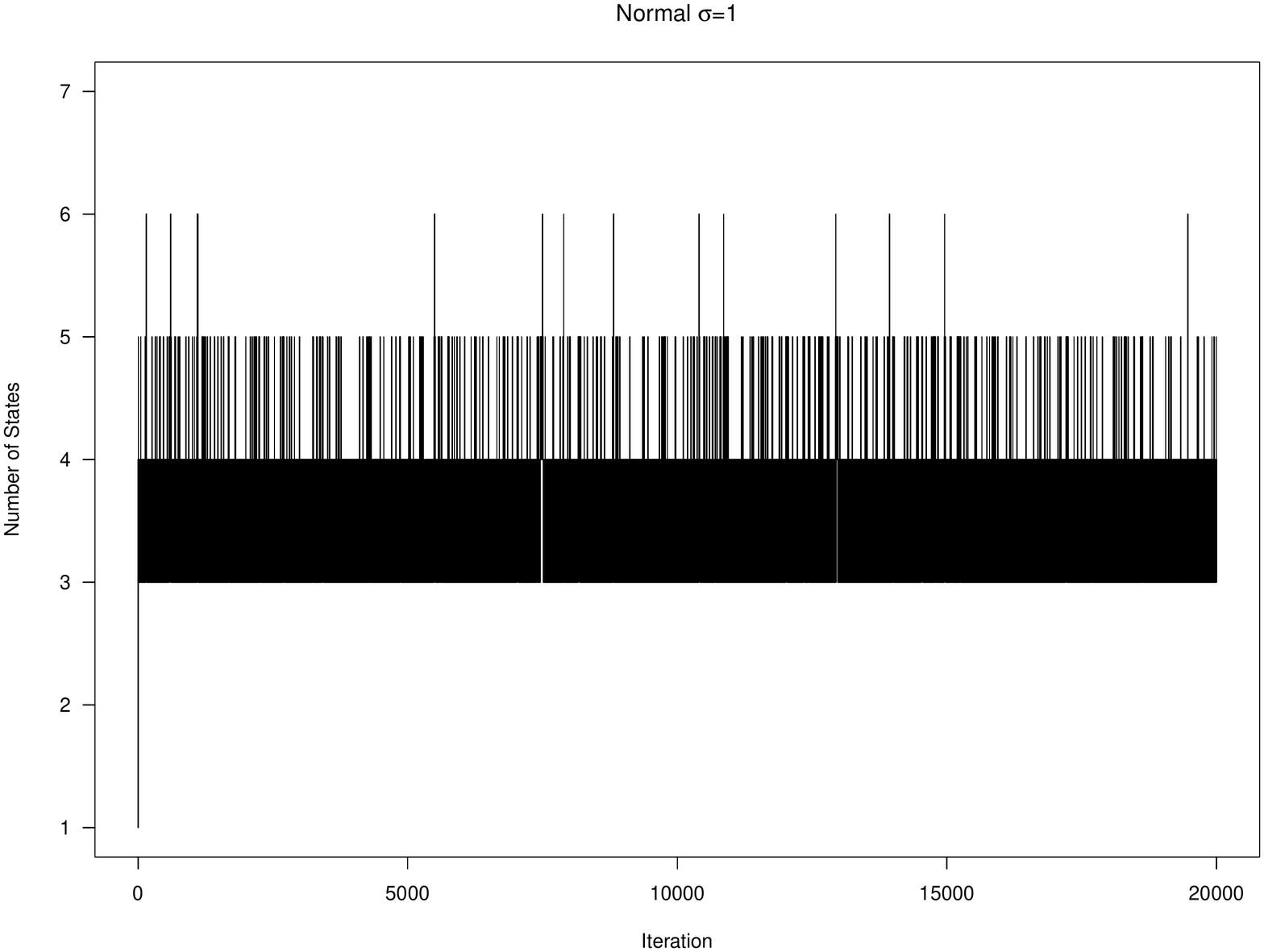}
	\end{minipage}
	\begin{minipage}[b]{0.45\textwidth}
		\includegraphics[width=\textwidth]{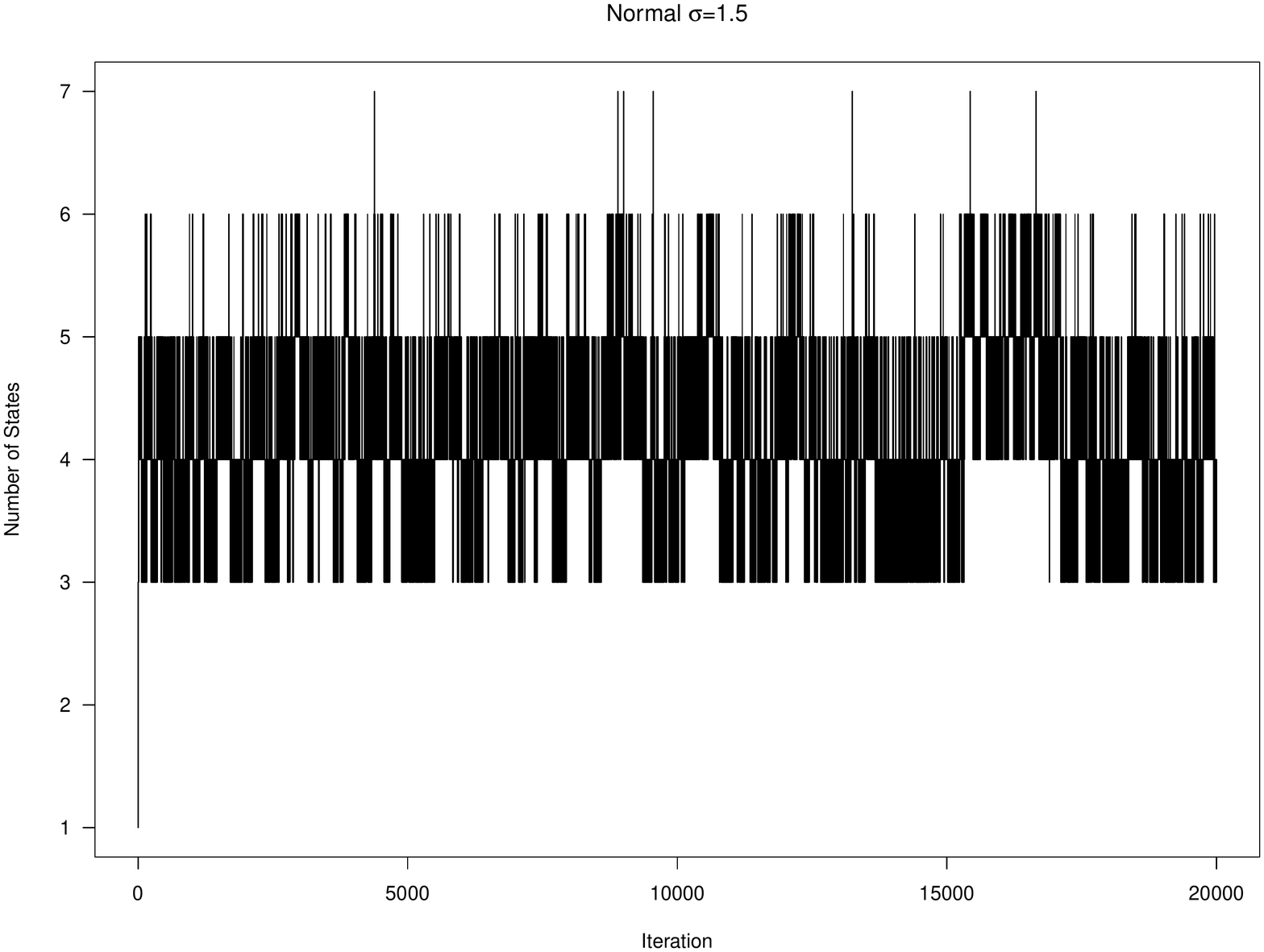}
	\end{minipage}
	\hfill
	\begin{minipage}[b]{0.45\textwidth}
		\includegraphics[width=\textwidth]{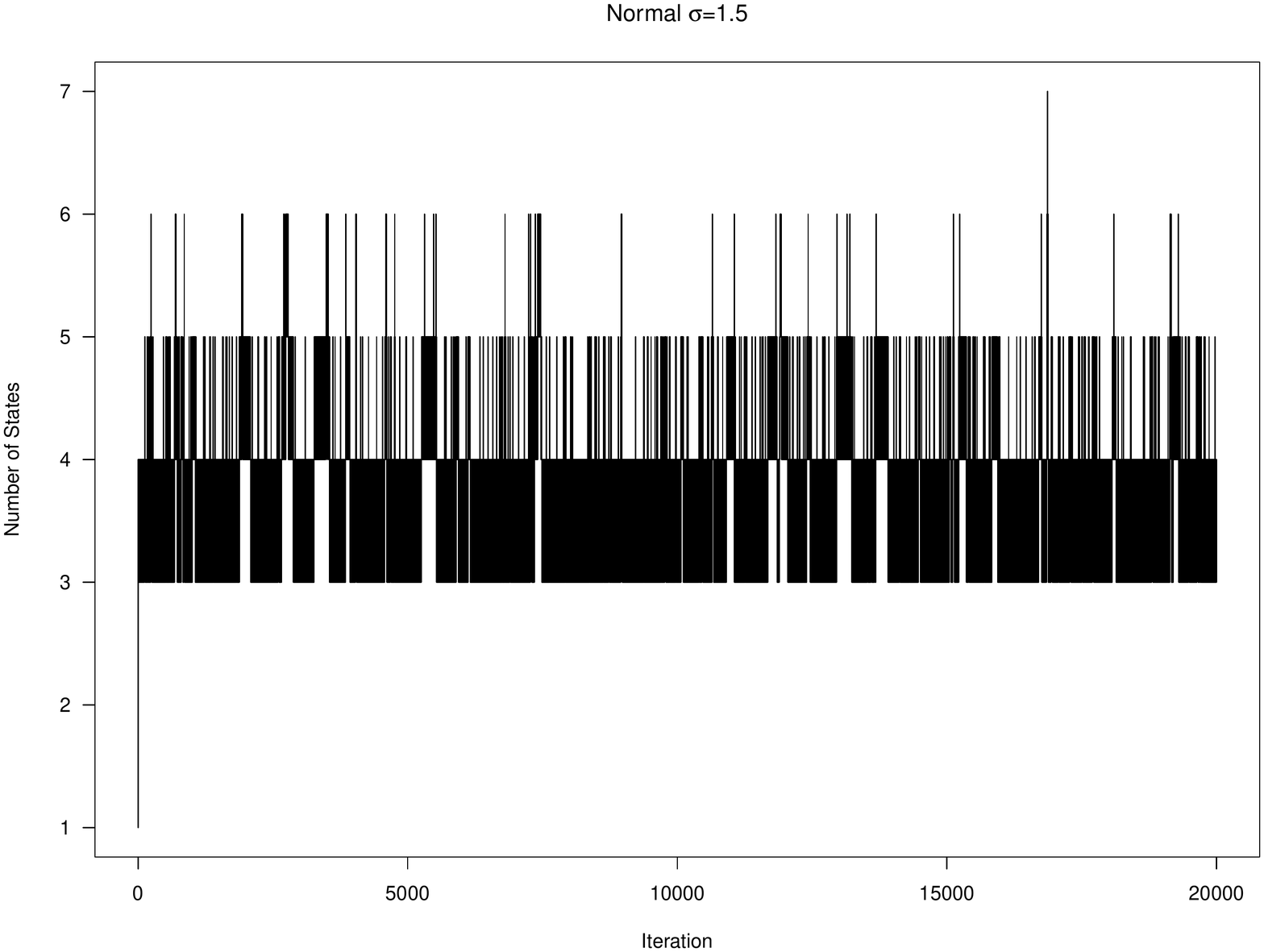}
	\end{minipage}
	\begin{minipage}[b]{0.45\textwidth}
		\includegraphics[width=\textwidth]{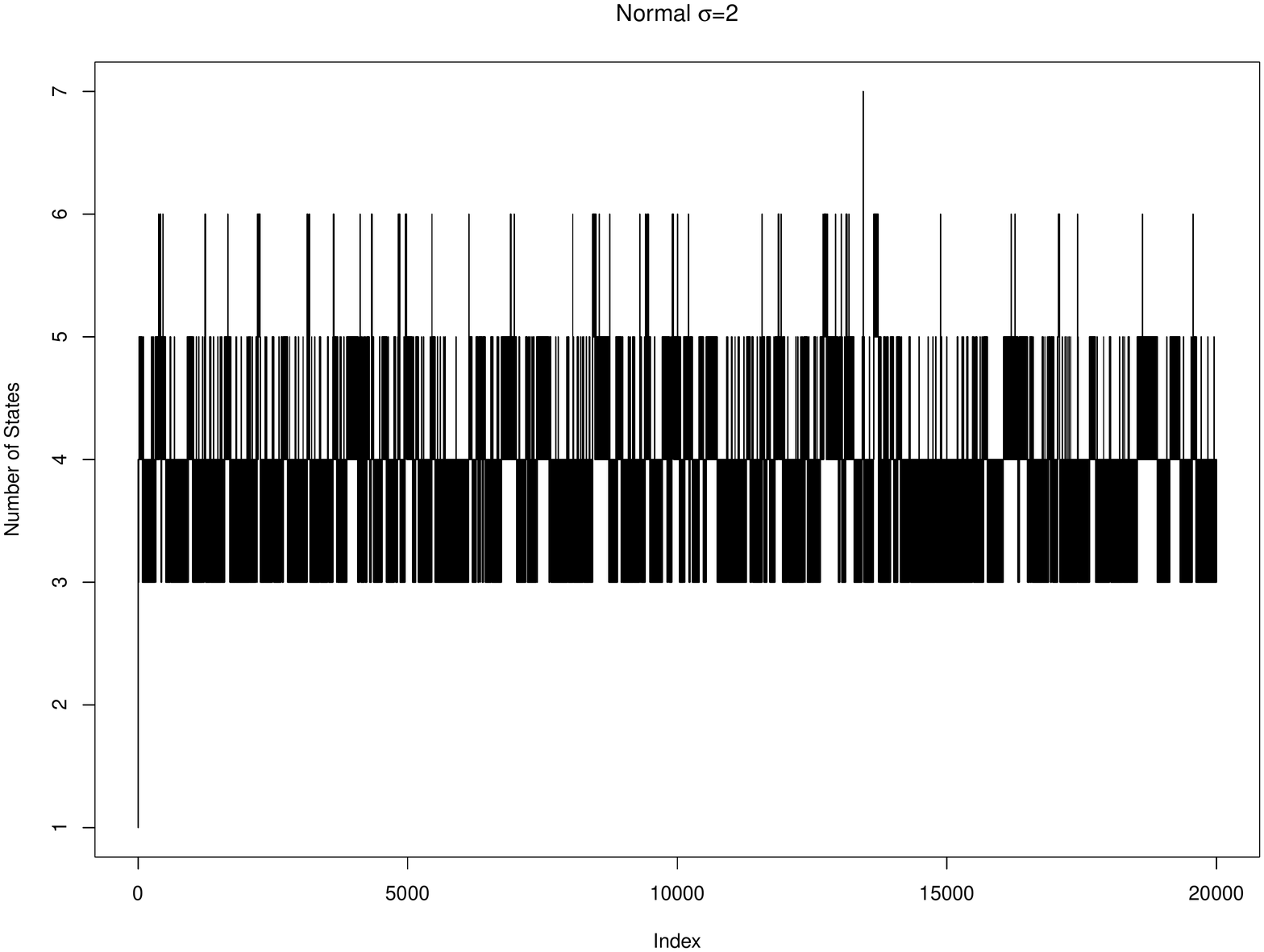}
	\end{minipage}
	\hfill
	\begin{minipage}[b]{0.45\textwidth}
		\includegraphics[width=\textwidth]{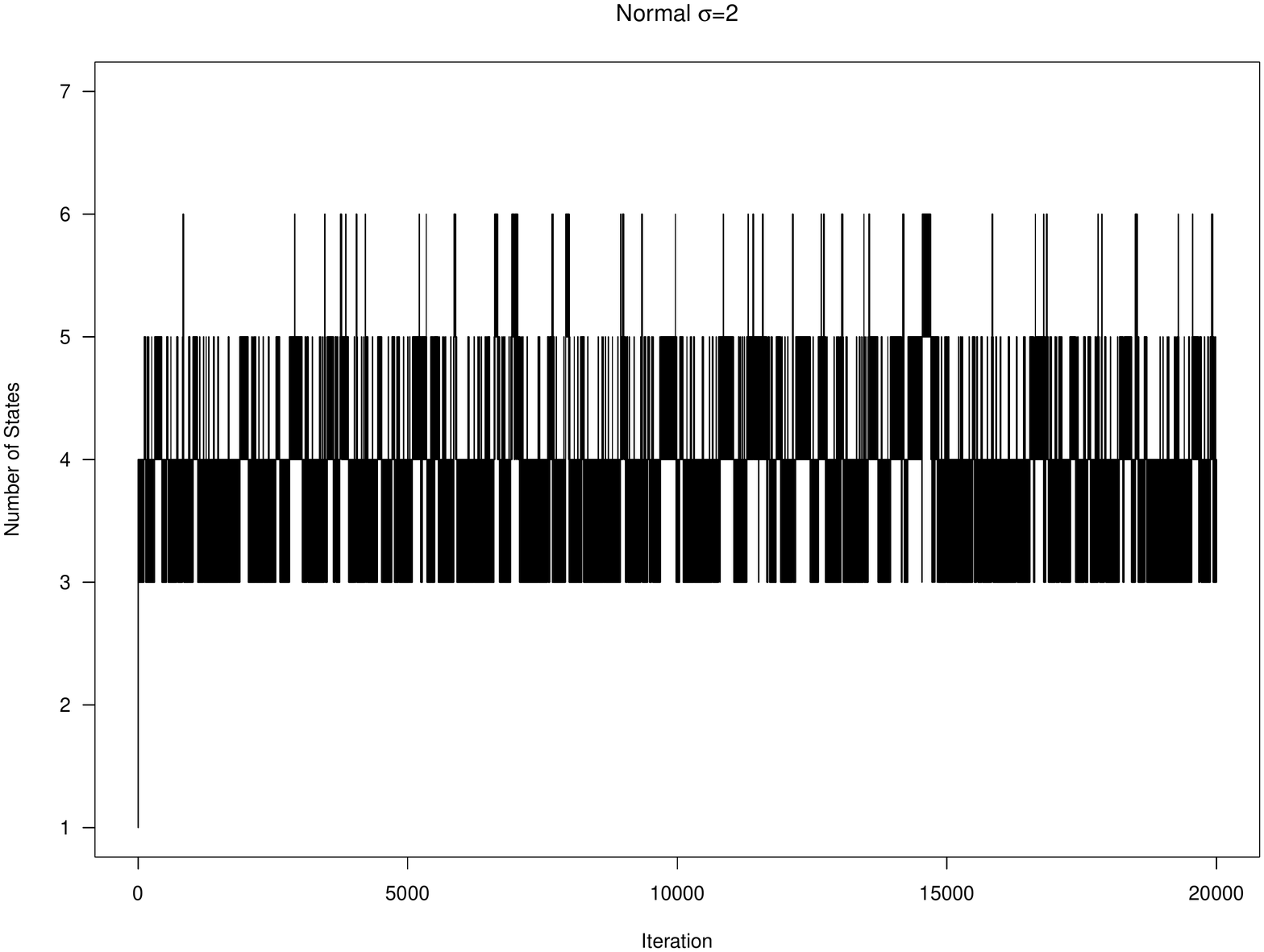}
	\end{minipage}
	\begin{minipage}[b]{0.45\textwidth}
		\includegraphics[width=\textwidth]{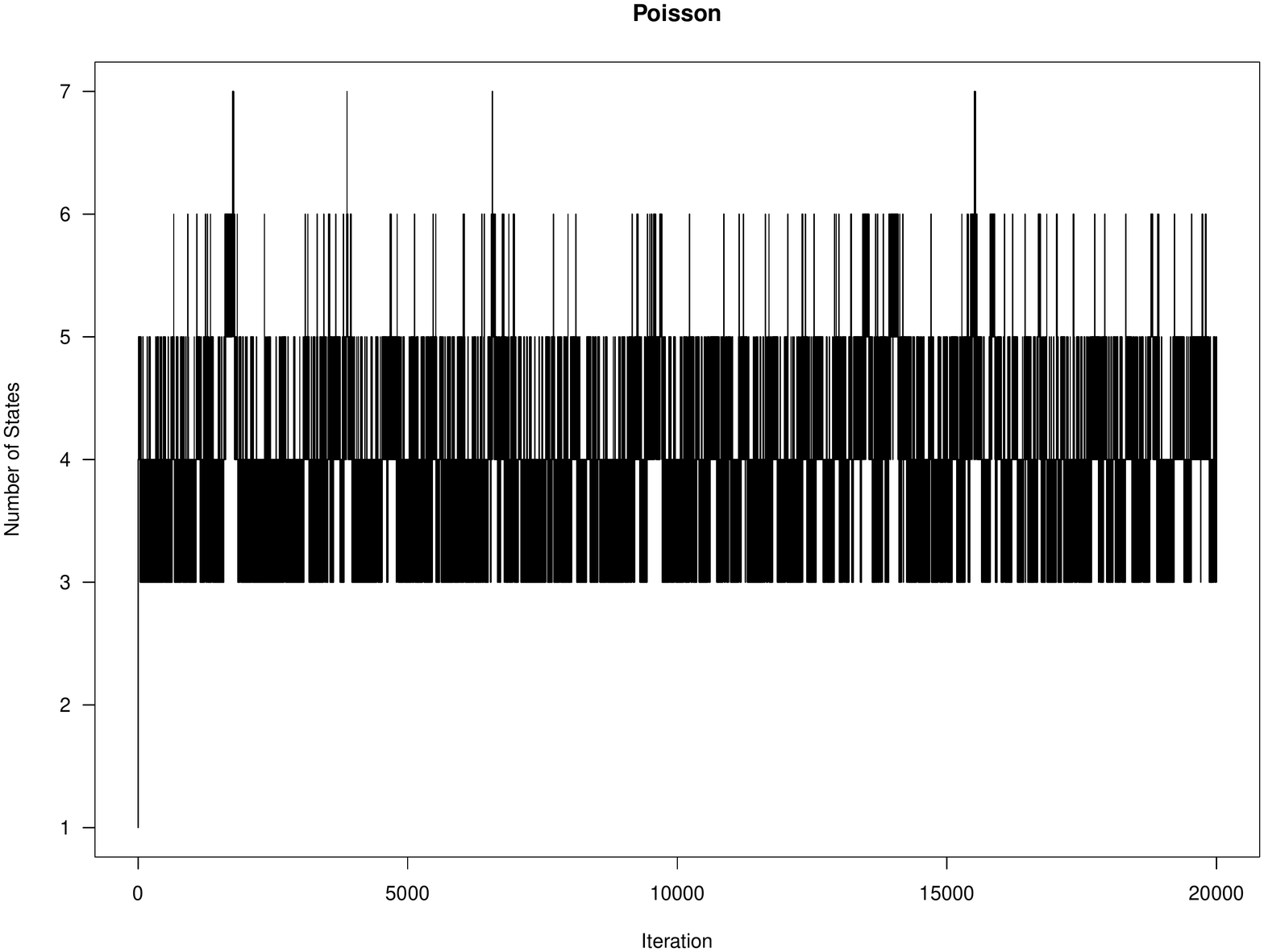}
	\end{minipage}
	\hfill
	\begin{minipage}[b]{0.45\textwidth}
		\includegraphics[width=\textwidth]{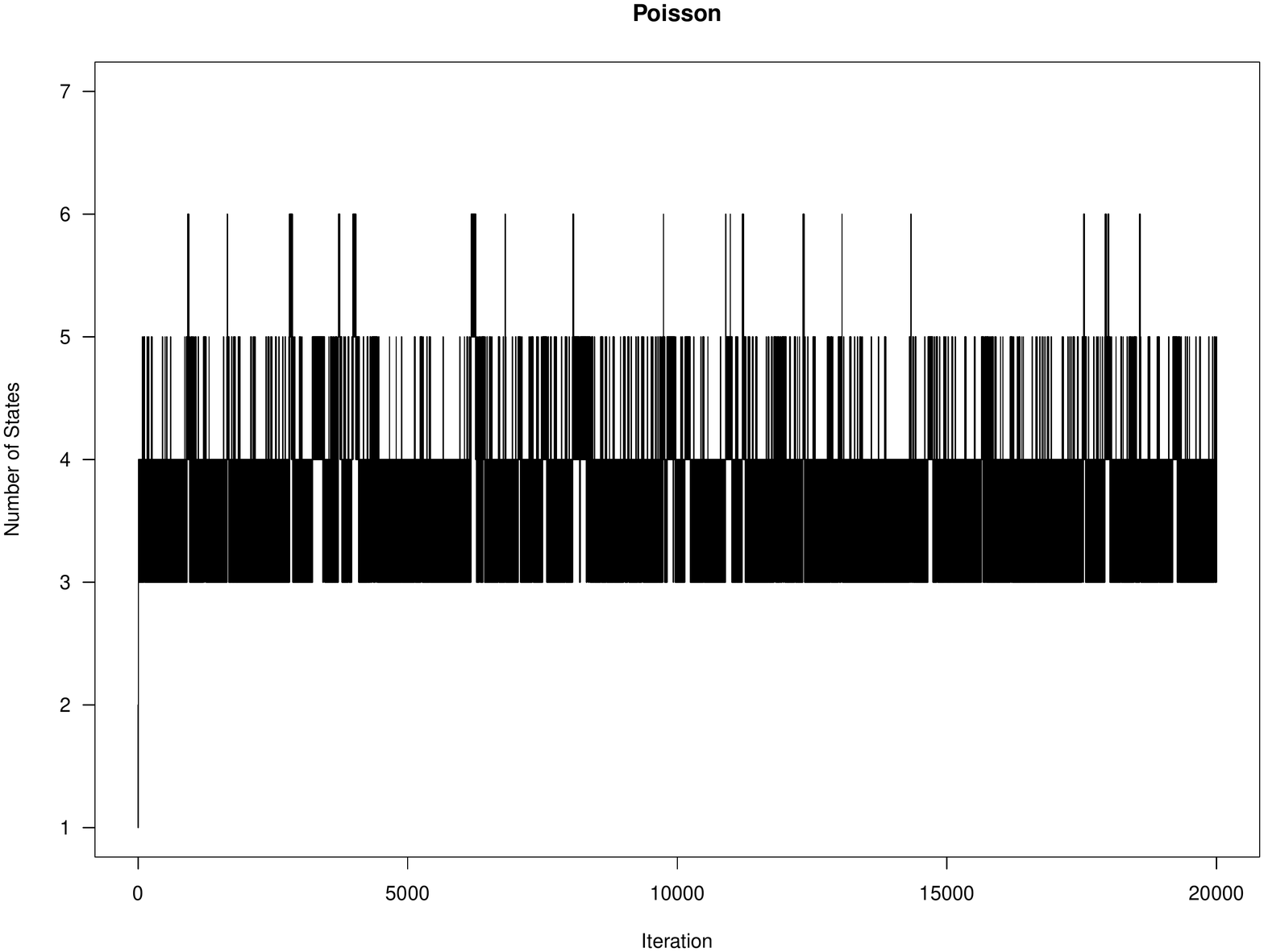}
	\end{minipage}
	
	\label{tracestatbdmcmc}
\end{figure}
The trace plots of the number of states for all the cases are shown in Figure \ref{tracestatbdmcmc} with the corresponding posterior distribution Table \ref{simex3}. As shown in the trace plots, the number of states change more frequently than reversible-jump MCMC. With a smaller $t_0$, the posterior distribution is more concentrated on three and four. However, in all cases, the posterior modes are four instead of three (the true number of states which the data were generating from); however, the posterior probabilities between three and four state models are close.

\section{Simulation: Trace Plots}
\begin{figure}[ht]
	\centering
	\caption{Example 1: Trace Plots for the Number of States of 20000 Iterations}
	\begin{minipage}[b]{0.45\textwidth}
		\includegraphics[width=\textwidth]{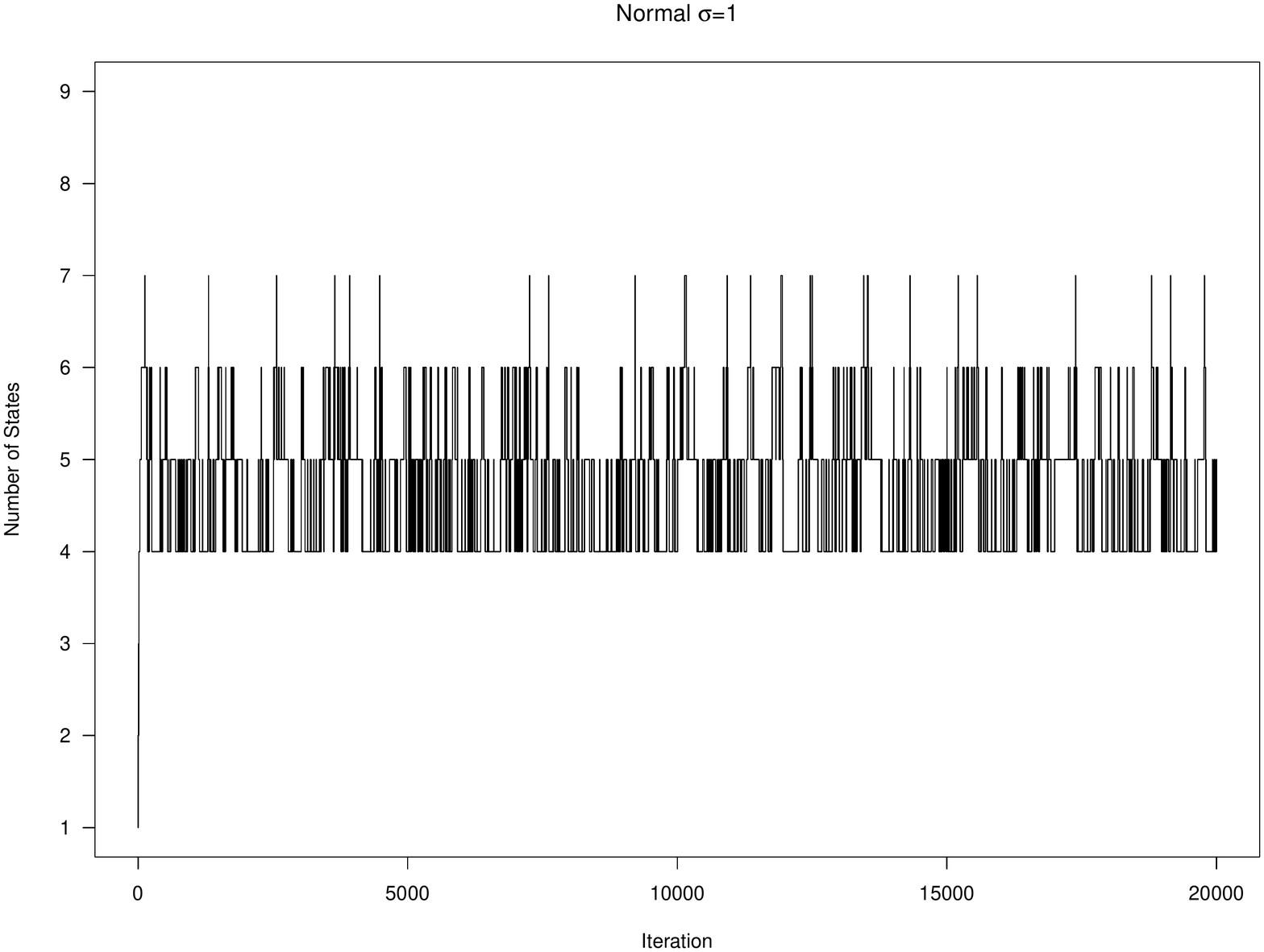}
	\end{minipage}
	\hfill
	\begin{minipage}[b]{0.45\textwidth}
		\includegraphics[width=\textwidth]{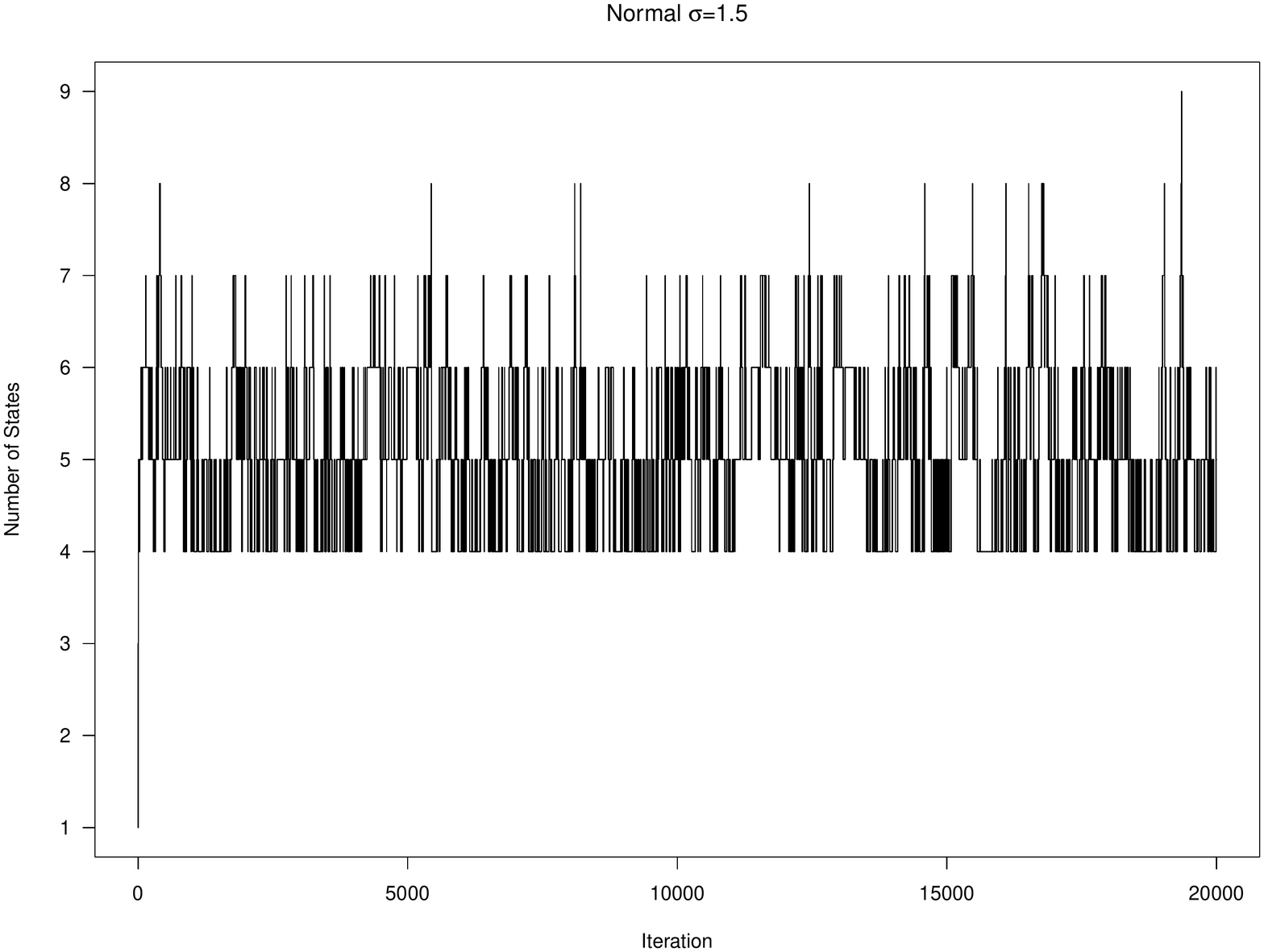}
	\end{minipage}
	\hfill
	\begin{minipage}[b]{0.45\textwidth}
		\includegraphics[width=\textwidth]{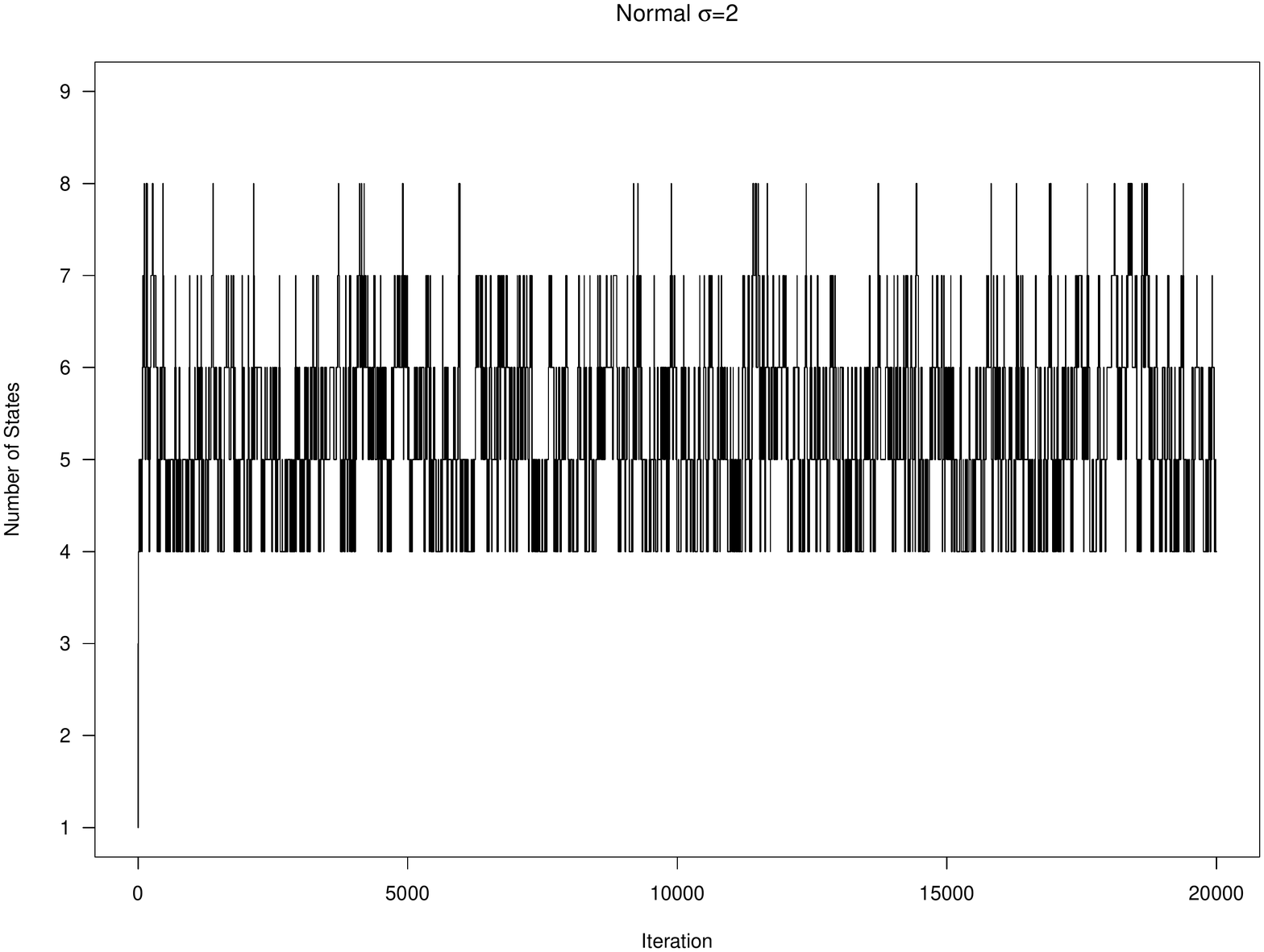}
	\end{minipage}
	\hfill
	\begin{minipage}[b]{0.45\textwidth}
		\includegraphics[width=\textwidth]{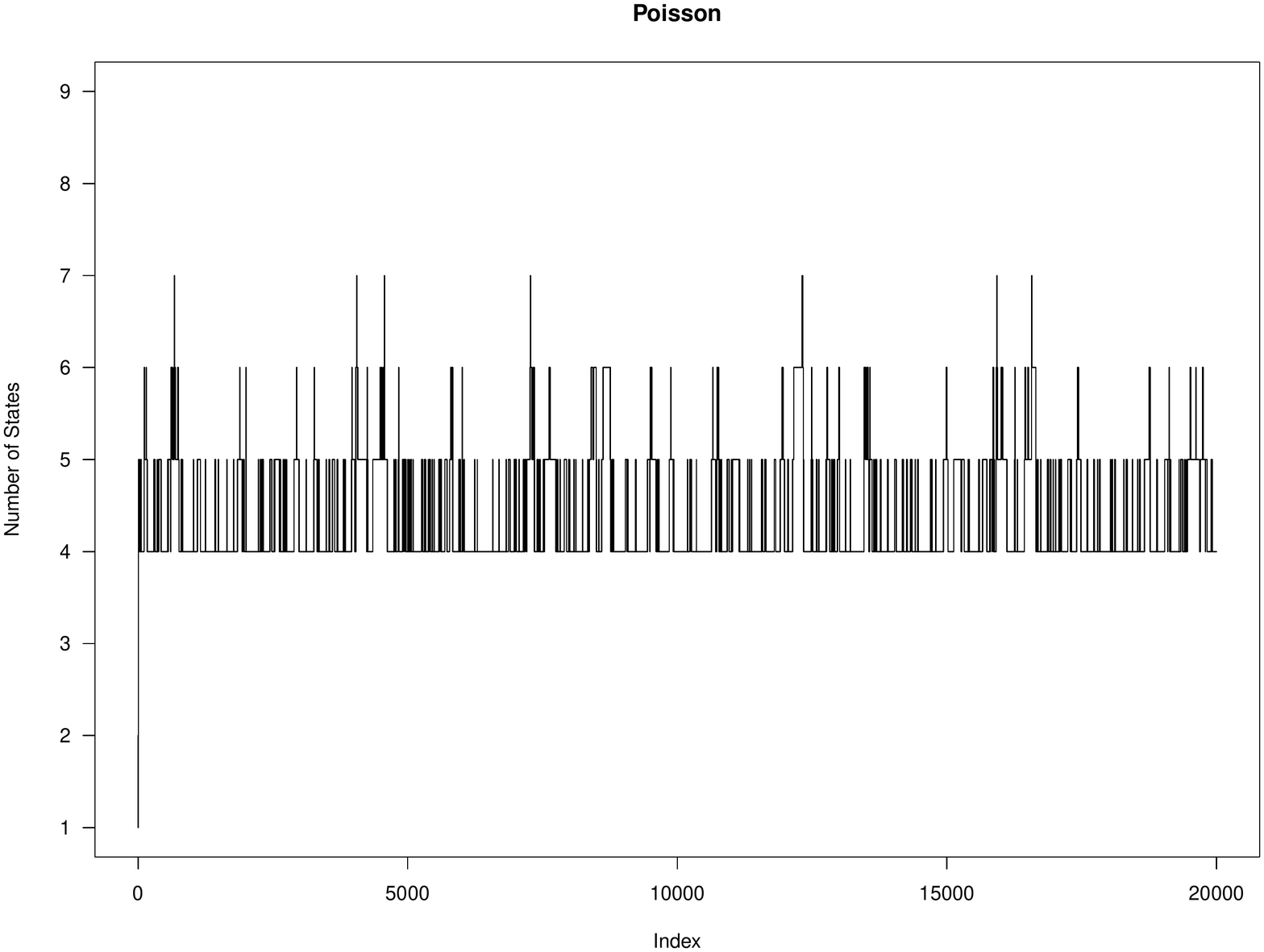}
	\end{minipage}
	\label{tracestates1}
\end{figure}

\begin{figure}[ht]
	\centering
	\caption{Example 3: Trace Plots for the Number of States of 20000 Iterations (Intercept only)}
	\begin{minipage}[b]{0.45\textwidth}
		\includegraphics[width=\textwidth]{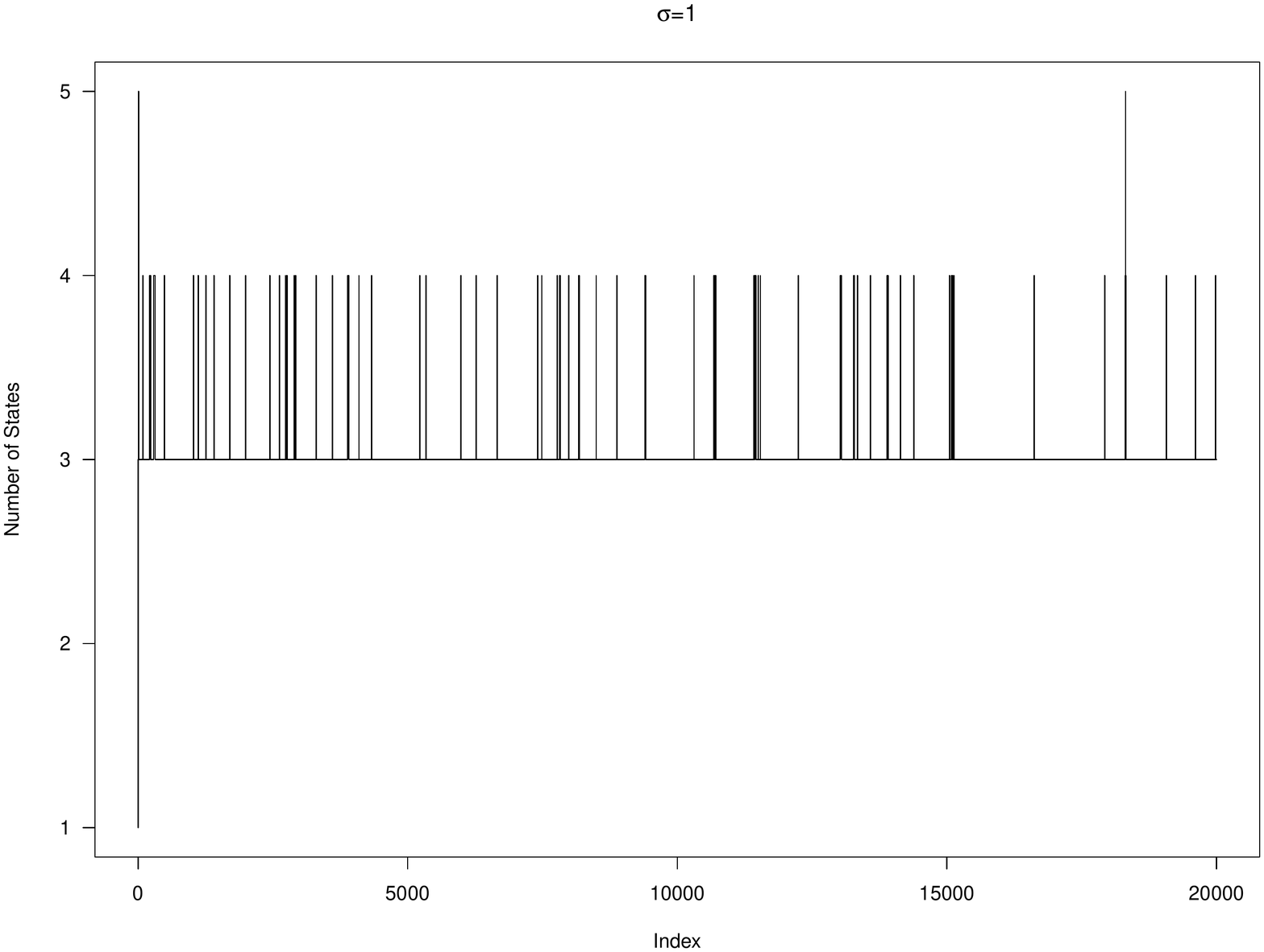}
	\end{minipage}
	\hfill
	\begin{minipage}[b]{0.45\textwidth}
		\includegraphics[width=\textwidth]{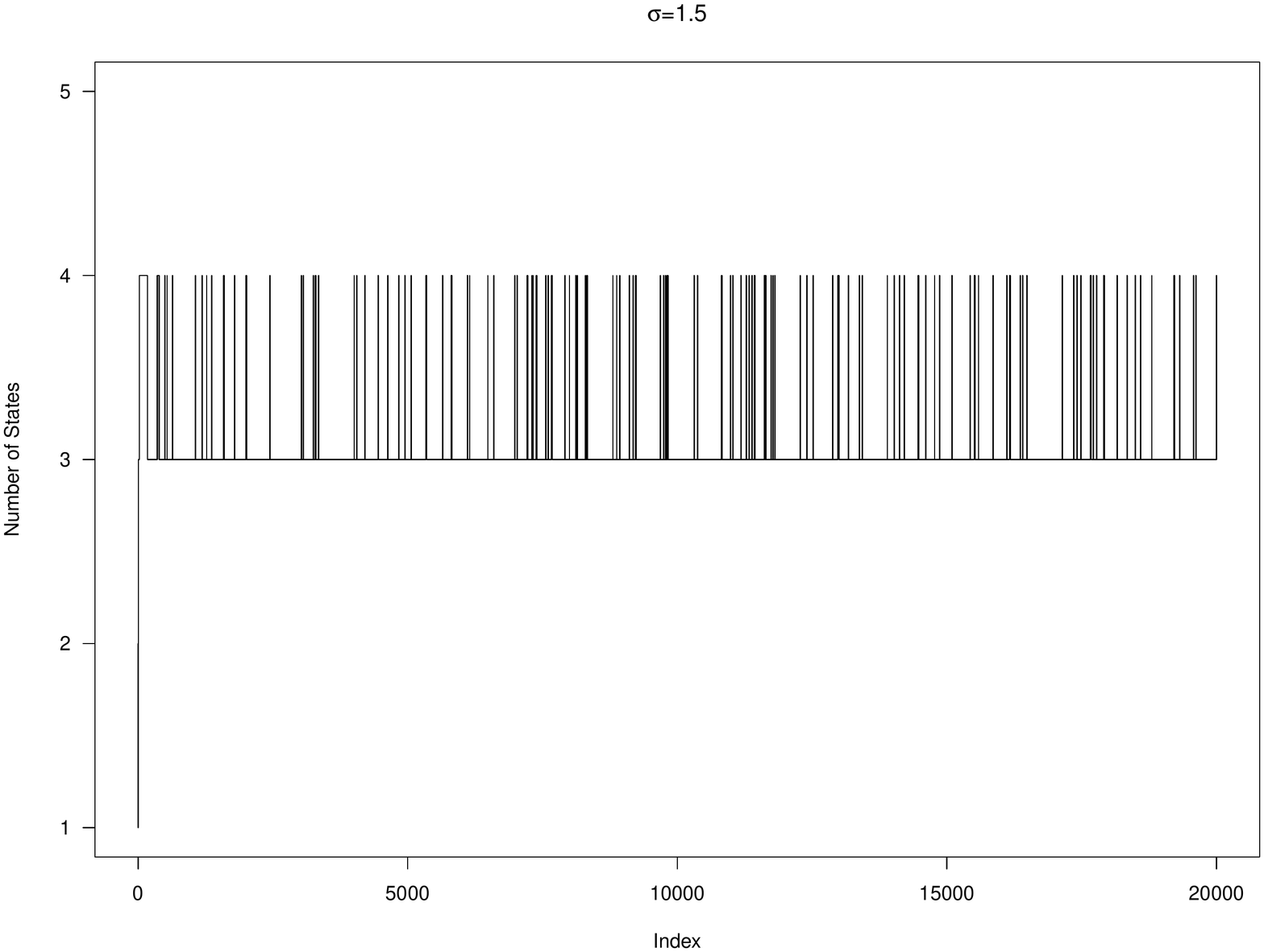}
	\end{minipage}
	\hfill
	\begin{minipage}[b]{0.45\textwidth}
		\includegraphics[width=\textwidth]{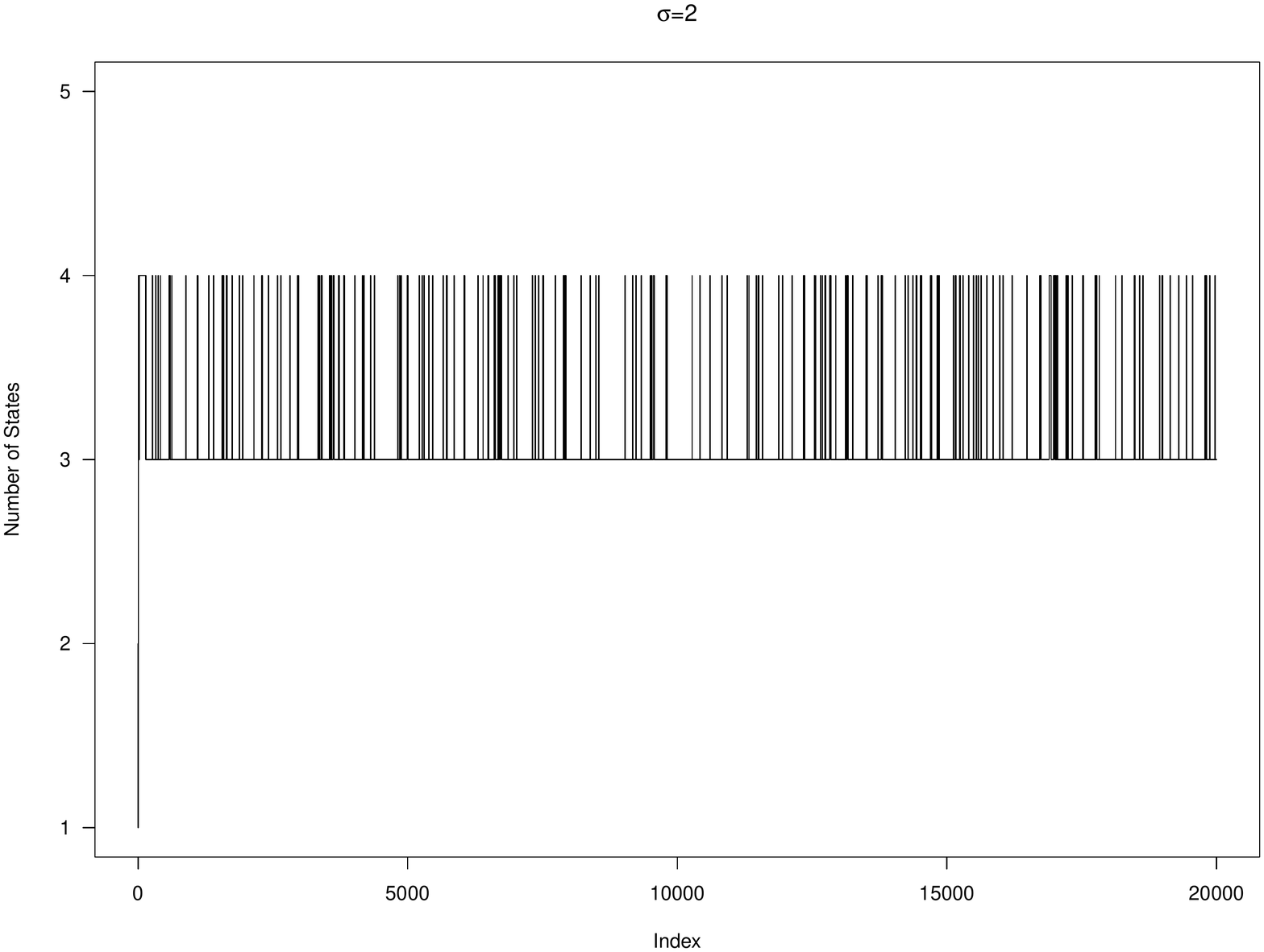}
	\end{minipage}
	\hfill
	\begin{minipage}[b]{0.45\textwidth}
		\includegraphics[width=\textwidth]{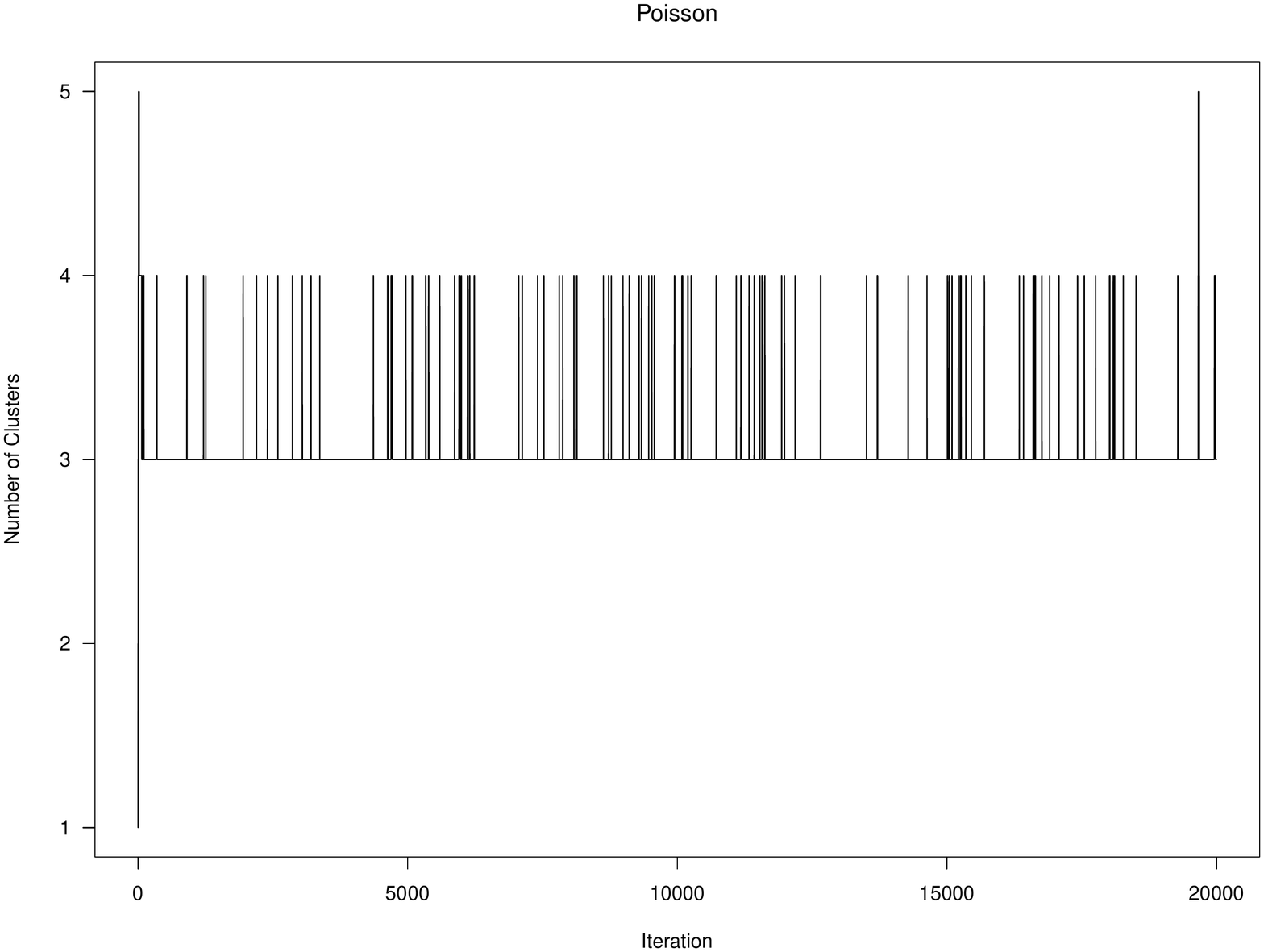}
	\end{minipage}
	\label{tracestates2}
\end{figure}

\begin{figure}[ht]
	\centering
	\caption{Example 4: Trace Plots for the Number of Clusters of 10000 Iterations.}
	\begin{minipage}[b]{0.45\textwidth}
		\includegraphics[width=\textwidth]{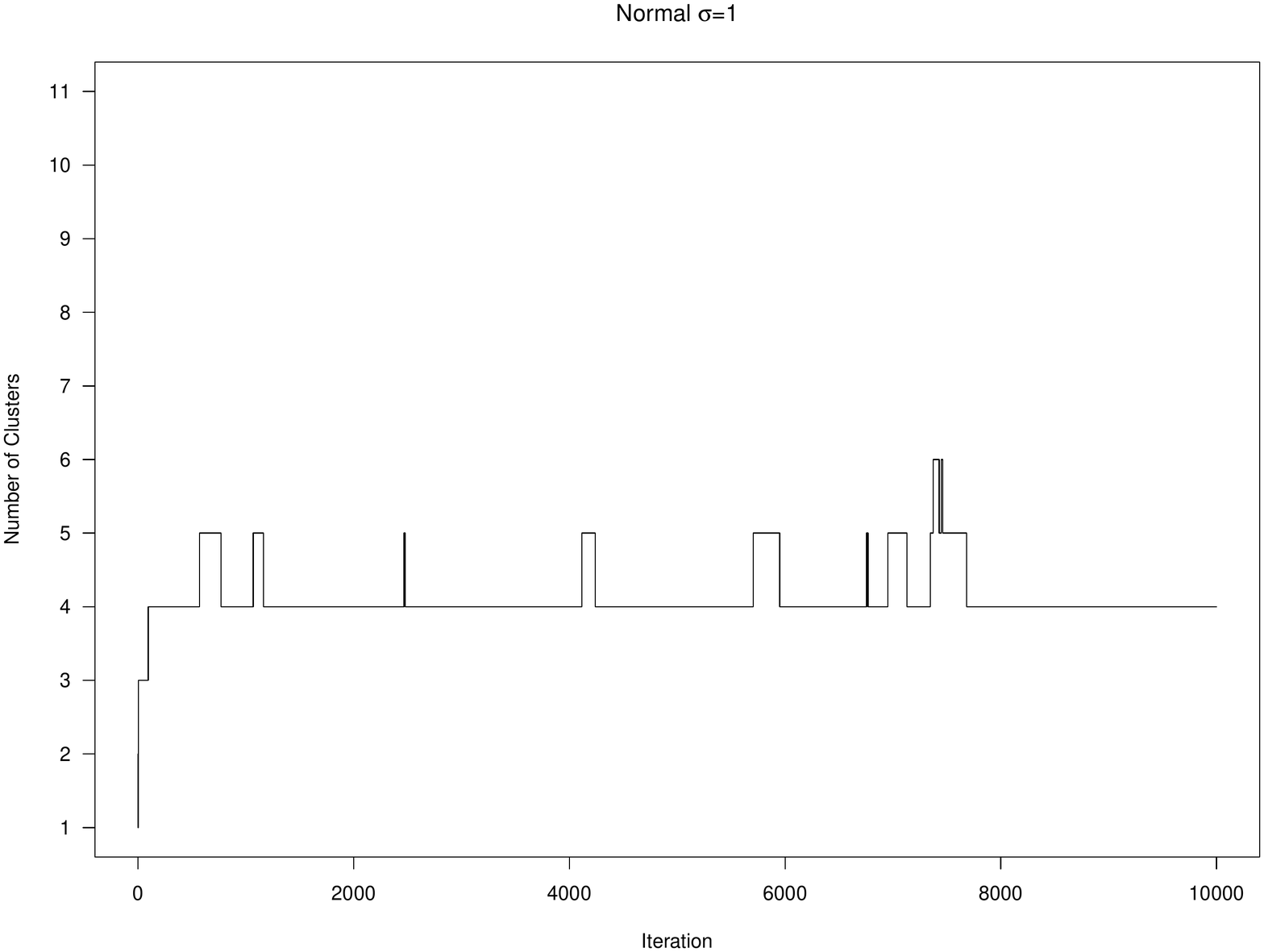}
	\end{minipage}
	\hfill
	\begin{minipage}[b]{0.45\textwidth}
		\includegraphics[width=\textwidth]{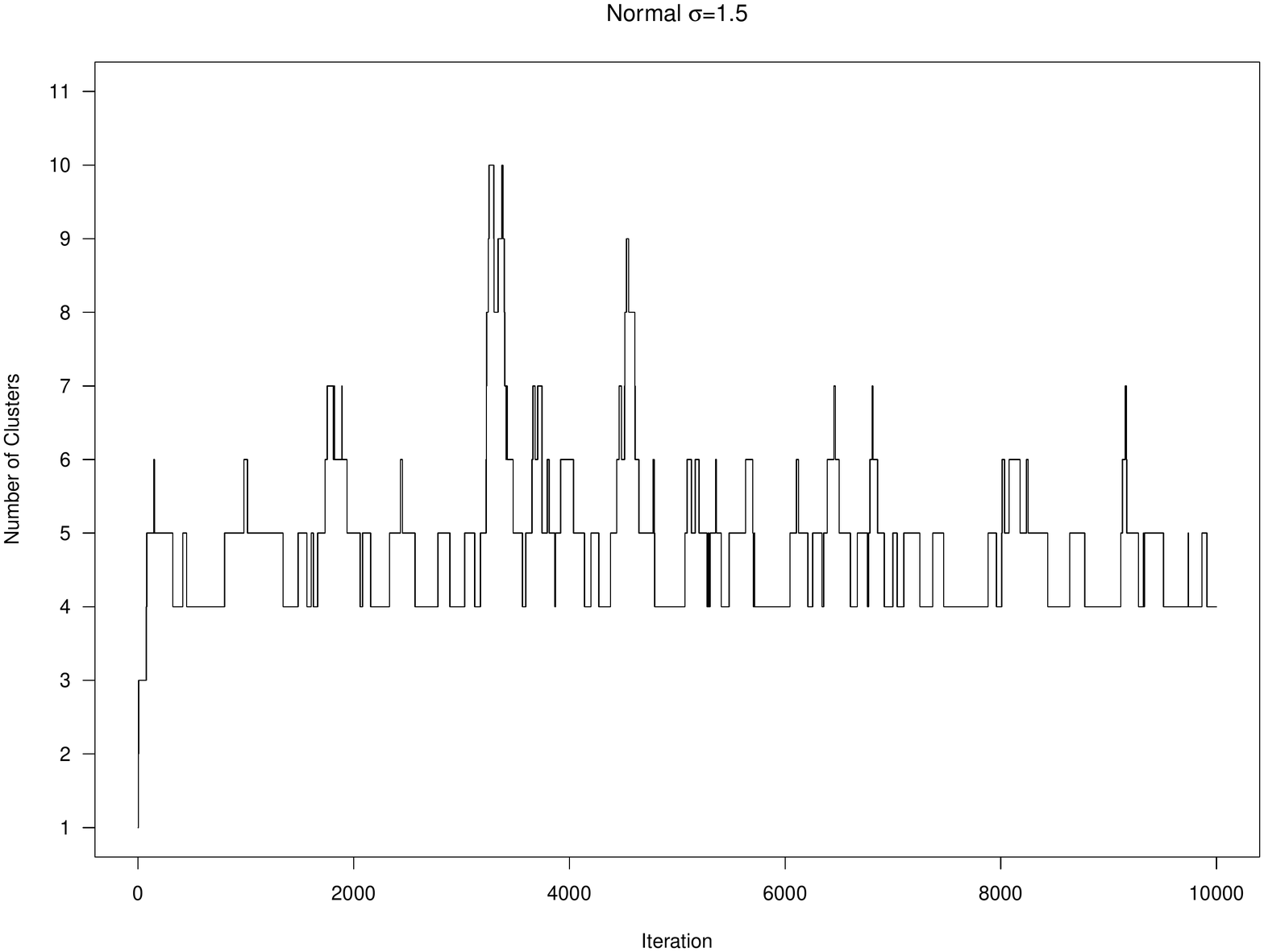}
	\end{minipage}
	\hfill
	\begin{minipage}[b]{0.45\textwidth}
		\includegraphics[width=\textwidth]{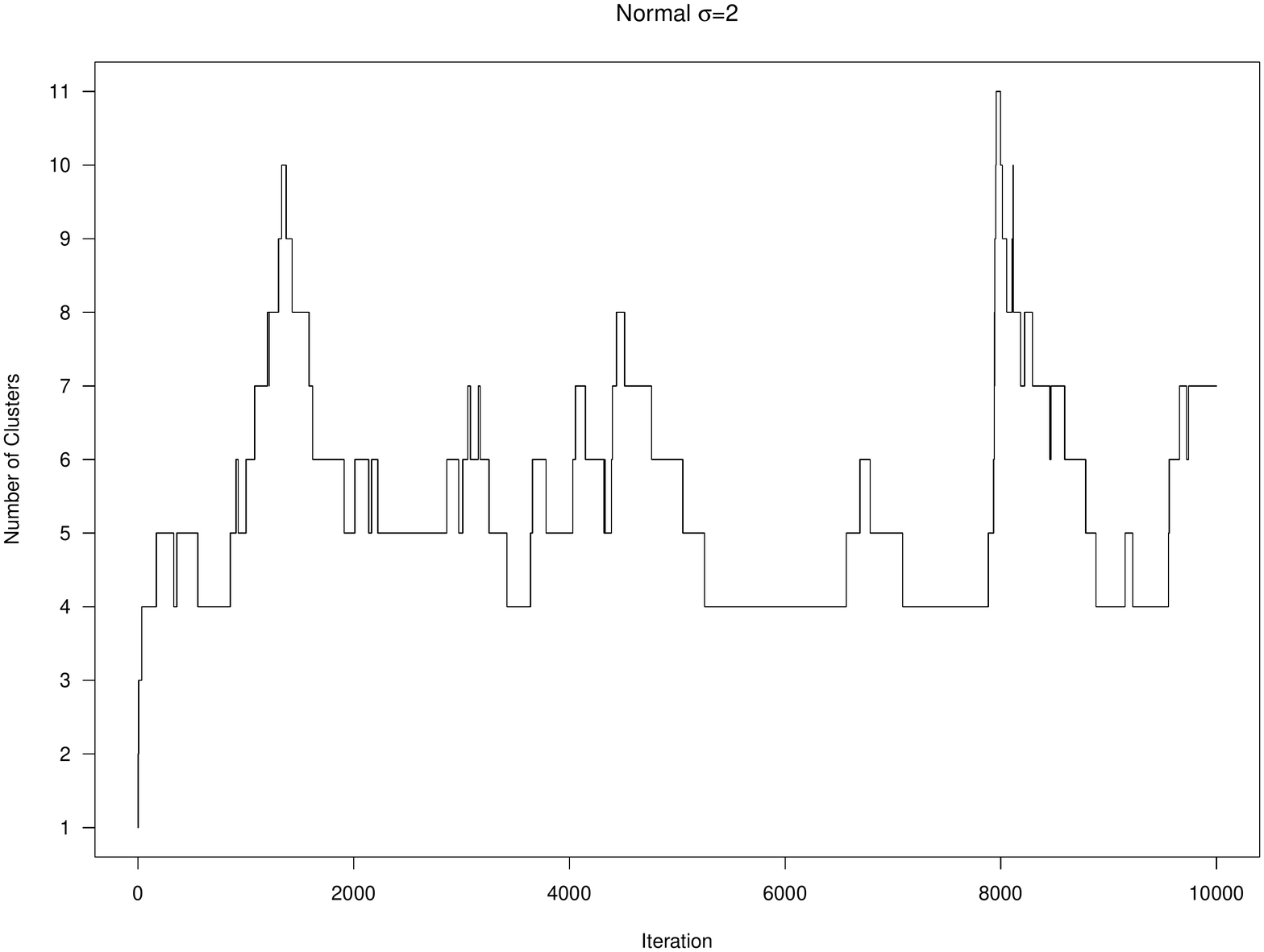}
	\end{minipage}
	\hfill
	\begin{minipage}[b]{0.45\textwidth}
		\includegraphics[width=\textwidth]{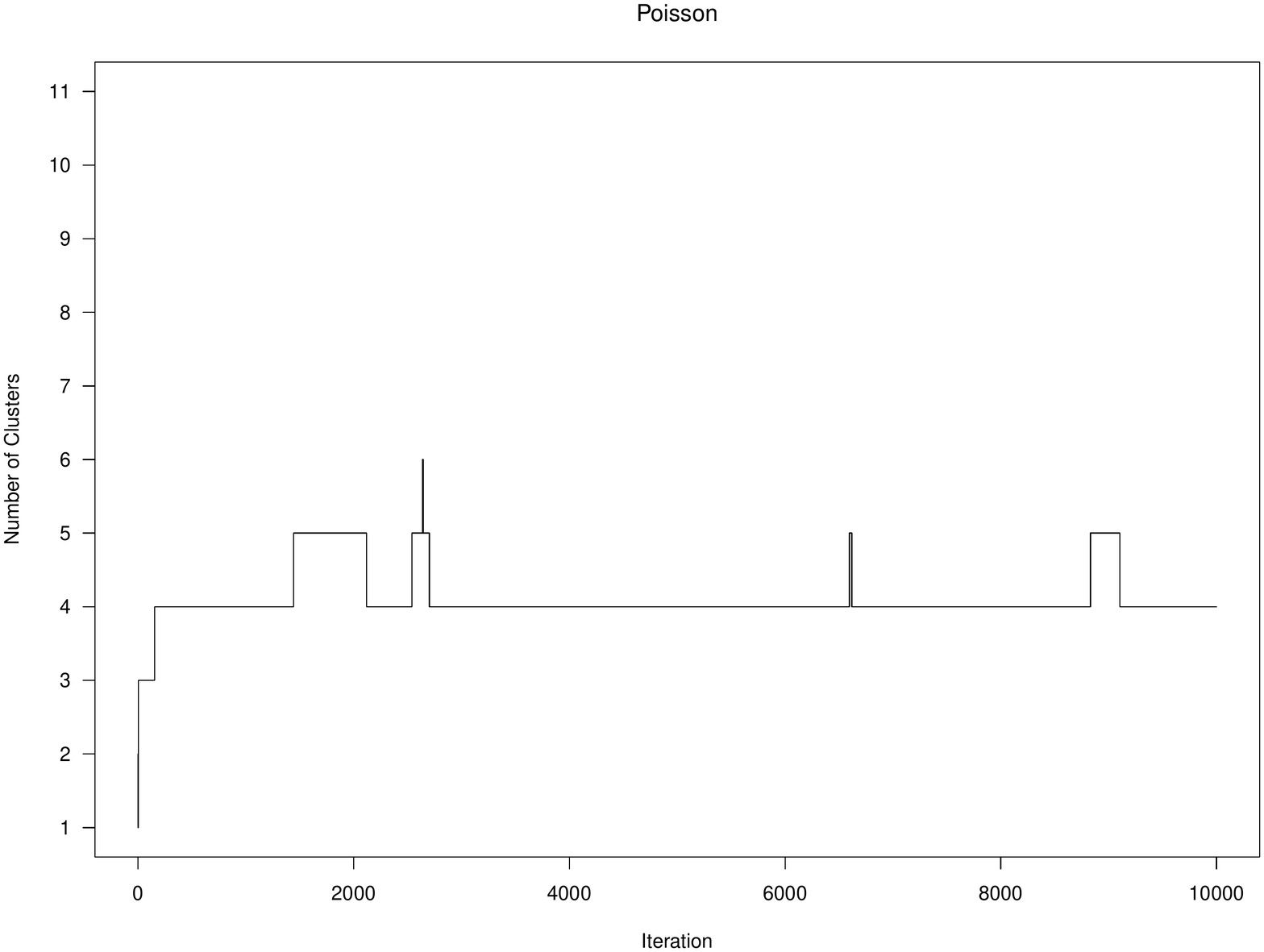}
	\end{minipage}
	\label{traceclustervar}
\end{figure}

\begin{figure}[ht]
	\centering
	\caption{Example 4: Trace plots of the number of states for Normal case $\sigma=1$ on four-cluster iterations.}
	\includegraphics[width=\textwidth]{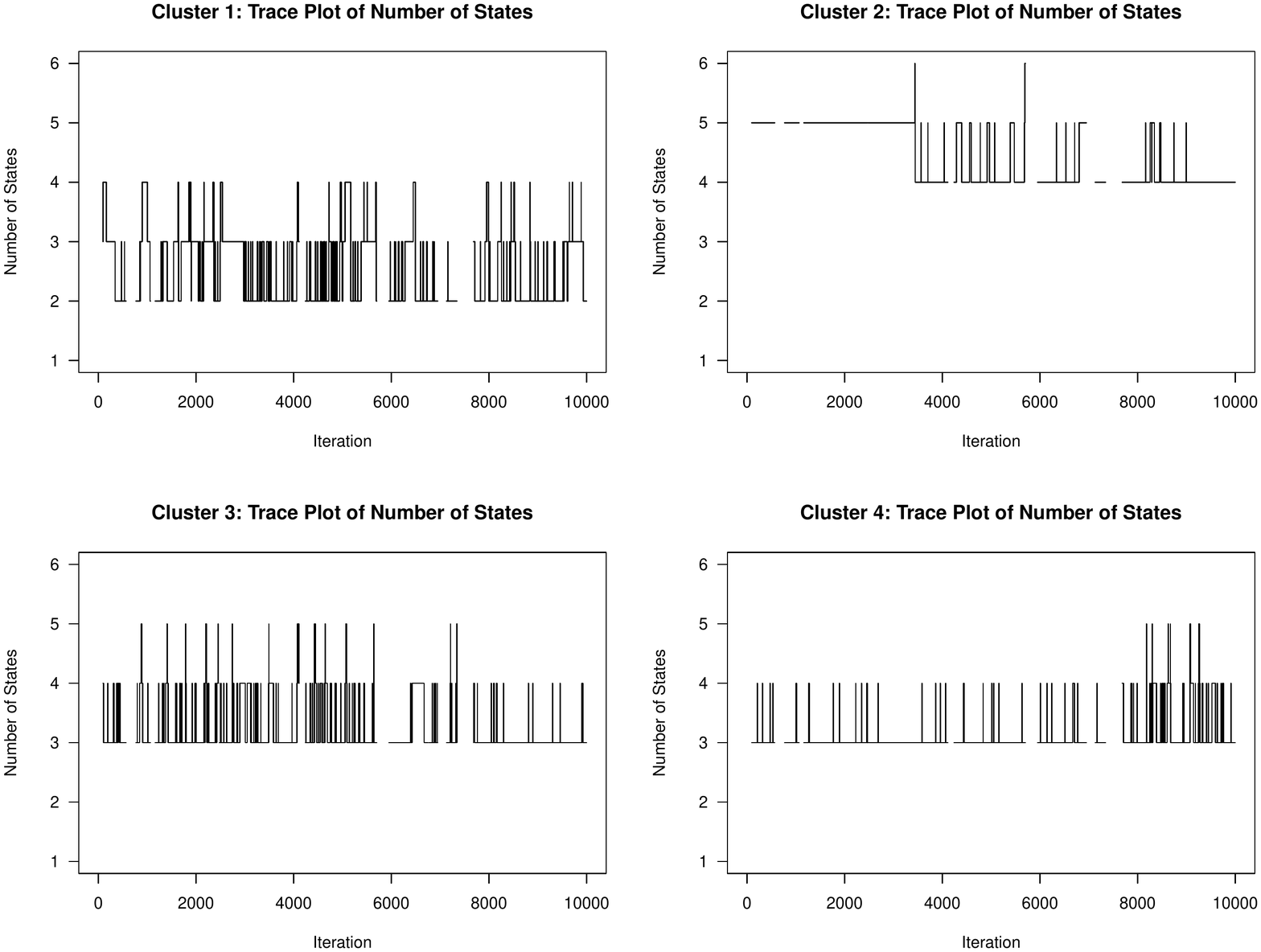}
	\label{clusterNor1}
\end{figure}

\begin{figure}[ht]
	\centering
	\caption{Example 4: Trace plots of the number of states for Normal case $\sigma=1.5$  on four-cluster iterations.}
	\includegraphics[width=\textwidth]{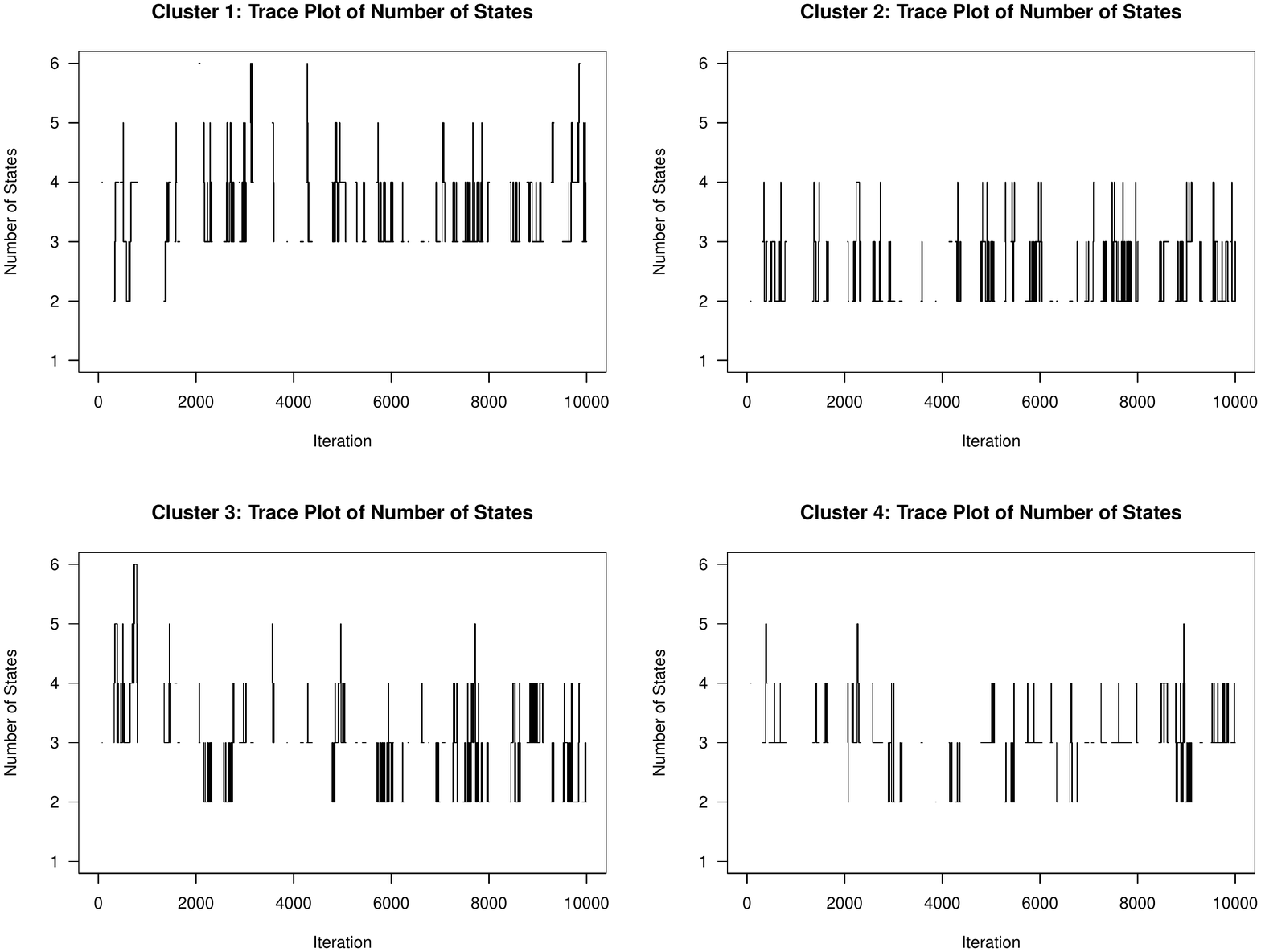}
	\label{clusterNor15}
\end{figure}

\begin{figure}[ht]
	\centering
	\caption{Example 4: Trace plots of the number of states for Normal case $\sigma=2$  on four-cluster iterations.}
	\includegraphics[width=\textwidth]{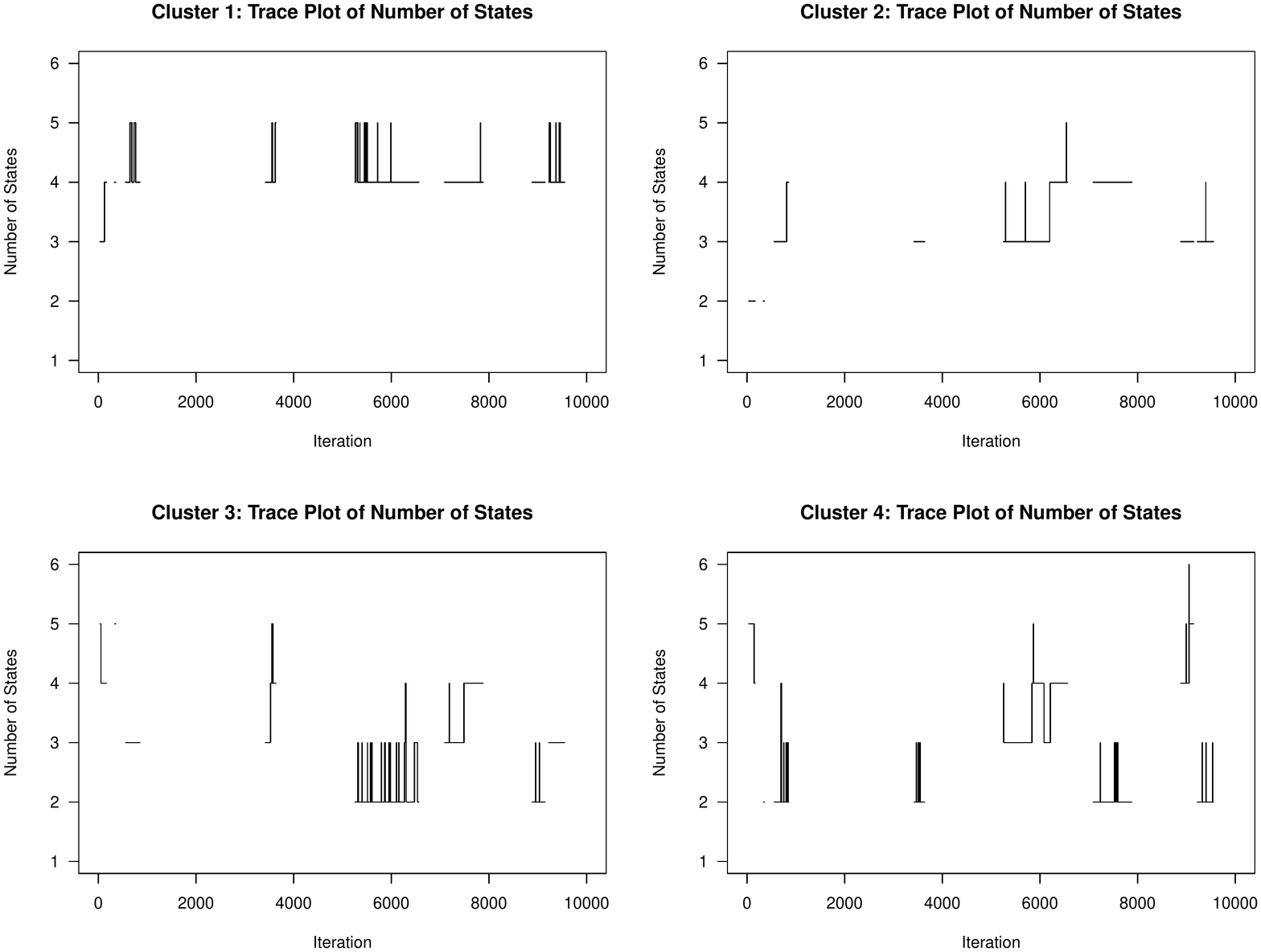}
	\label{clusterNor2}
\end{figure}

\begin{figure}[ht]
	\centering
	\caption{Example 4: Trace plots of the number of states for Poisson case on four-cluster iterations.}
	\includegraphics[width=\textwidth]{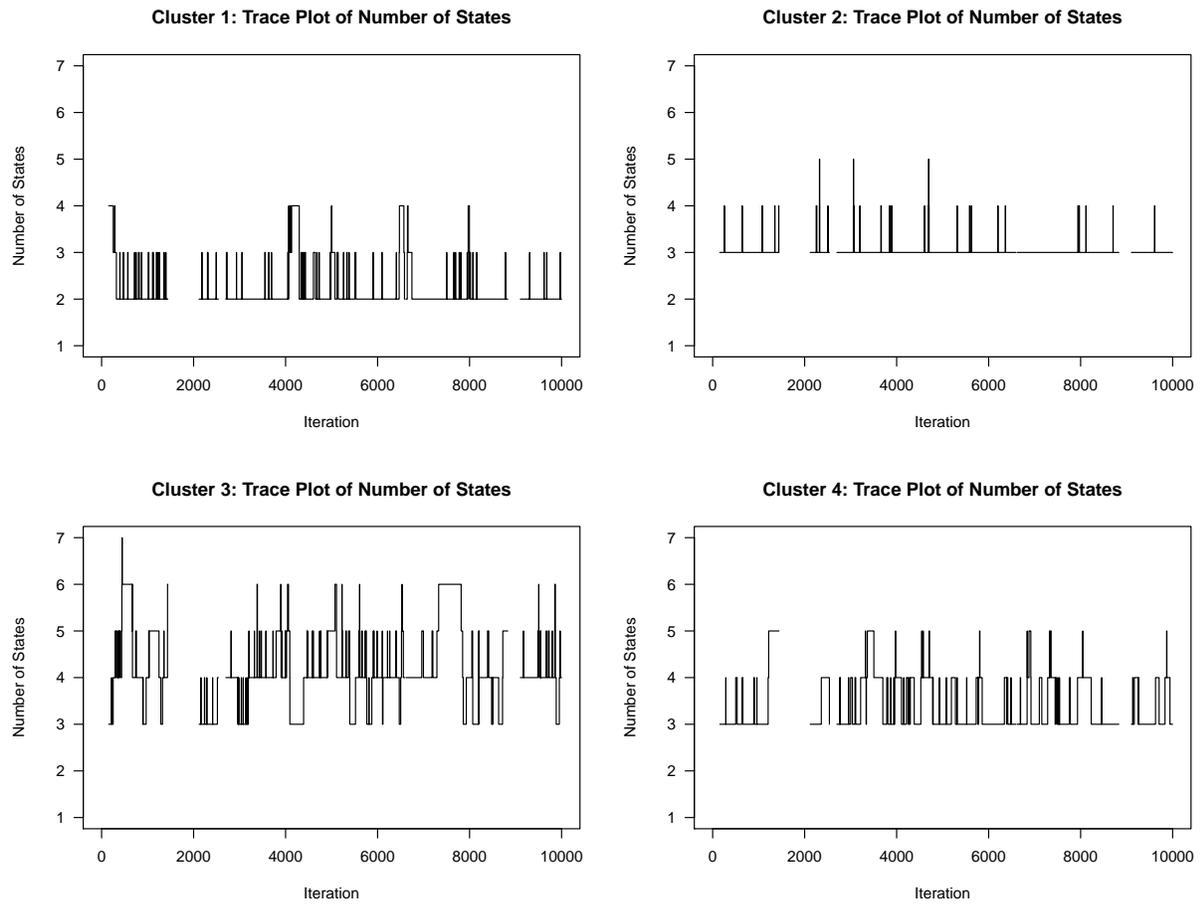}
	\label{clusterpois}
\end{figure}

\end{document}